\PassOptionsToPackage{unicode}{hyperref}
\PassOptionsToPackage{hyphens}{url}
\documentclass[
]{article}
\usepackage{amsmath,amssymb}
\usepackage{iftex}
\ifPDFTeX
  \usepackage[T1]{fontenc}
  \usepackage[utf8]{inputenc}
  \usepackage{textcomp} % provide euro and other symbols
\else % if luatex or xetex
  \usepackage{unicode-math} % this also loads fontspec
  \defaultfontfeatures{Scale=MatchLowercase}
  \defaultfontfeatures[\rmfamily]{Ligatures=TeX,Scale=1}
\fi
\usepackage{lmodern}
\ifPDFTeX\else
\fi
\IfFileExists{upquote.sty}{\usepackage{upquote}}{}
\IfFileExists{microtype.sty}{% use microtype if available
  \usepackage[]{microtype}
  \UseMicrotypeSet[protrusion]{basicmath} % disable protrusion for tt fonts
}{}
\makeatletter
\@ifundefined{KOMAClassName}{% if non-KOMA class
  \IfFileExists{parskip.sty}{%
    \usepackage{parskip}
  }{% else
    \setlength{\parindent}{0pt}
    \setlength{\parskip}{6pt plus 2pt minus 1pt}}
}{% if KOMA class
  \KOMAoptions{parskip=half}}
\makeatother
\usepackage{xcolor}
\usepackage[margin=25.4mm]{geometry}
\usepackage{graphicx}
\makeatletter
\def\maxwidth{\ifdim\Gin@nat@width>\linewidth\linewidth\else\Gin@nat@width\fi}
\def\maxheight{\ifdim\Gin@nat@height>\textheight\textheight\else\Gin@nat@height\fi}
\makeatother
\setkeys{Gin}{width=\maxwidth,height=\maxheight,keepaspectratio}
\makeatletter
\def\fps@figure{htbp}
\makeatother
\newlength{\cslhangindent}
\newlength{\csllabelwidth}
\newlength{\cslentryspacingunit} % times entry-spacing
\newenvironment{CSLReferences}[2] % #1 hanging-ident, #2 entry spacing
 {% don't indent paragraphs
  \setlength{\parindent}{0pt}
  \ifodd #1
  \let\oldpar\par
  \def\par{\hangindent=\cslhangindent\oldpar}
  \fi
  \setlength{\parskip}{#2\cslentryspacingunit}
 }%
 {}
\usepackage{calc}

\usepackage{setspace}
\usepackage{parskip}
\usepackage{booktabs}
\usepackage{multirow}
\usepackage{makecell}
\usepackage[utf8]{inputenc}
\usepackage{amsmath}
\usepackage{lineno}
\usepackage{subcaption}
\usepackage{graphicx}
\usepackage[font=small,skip=0pt]{caption}
\usepackage{float} % here for H placement parameter
\usepackage[title]{appendix}
\usepackage{xcolor}
\usepackage{soul}
\ifLuaTeX
  \usepackage{selnolig}  % disable illegal ligatures
\fi
\IfFileExists{bookmark.sty}{\usepackage{bookmark}}{\usepackage{hyperref}}
\IfFileExists{xurl.sty}{\usepackage{xurl}}{} % add URL line breaks if available
\hypersetup{
  pdftitle={Model-assisted estimation of domain totals, areas, and densities in two-stage sample survey designs},
  pdfauthor={Hans-Erik Andersen1,*; Göran Ståhl2; Bruce D. Cook3; Douglas C. Morton3; Andrew O. Finley4},
  hidelinks,
  pdfcreator={LaTeX via pandoc}}

\begin{document}

\begin{center}
{\LARGE Model-assisted estimation of domain totals, areas, and densities
in two-stage sample survey designs}
\vspace{0.5cm}

Hans-Erik Andersen$^1$, Göran Ståhl$^2$, Bruce D. Cook$^3$, Douglas C. Morton$^3$, Andrew O. Finley$^{4}$

\vspace{5mm}
{\small
  $^1$USDA Forest Service, Pacific Northwest Research Station, Seattle, WA, USA\\
  $^2$Department of Forest Resource Management, Swedish University of Agricultural Sciences, Umeå, Sweden\\
  $^3$Goddard Space Flight Center, National Aeronautics and Space Administration, Greenbelt, MD, USA\\
  $^4$Departments of Forestry and Statistics \& Probability, Michigan State University, East Lansing, MI, USA\\
  
  \vspace{0.5cm}
  \textbf{Corresponding Author}: Hans-Erik Andersen, hans.andersen@usda.gov
  }
\end{center}

\section*{Abstract}
Model-assisted, two-stage forest survey sampling designs provide a means
to combine airborne remote sensing data, collected in a sampling mode,
with field plot data to increase the precision of national forest
inventory estimates, while maintaining important properties of
design-based inventories, such as unbiased estimation and quantification
of uncertainty. In this study, we present a comprehensive set of
model-assisted estimators for domain-level attributes in a two-stage
sampling design, including new estimators for densities, and compare the
performance of these estimators with standard poststratified estimators.
Simulation was used to assess the statistical properties (bias,
variability) of these estimators, with both simple random and systematic
sampling configurations, and indicated that 1) all estimators were
generally unbiased. and 2) the use of lidar in a sampling mode increased
the precision of the estimators at all assessed field sampling
intensities, with particularly marked increases in precision at lower
field sampling intensities. Variance estimators are generally unbiased
for model-assisted estimators without poststratification, while
model-assisted estimators with poststratification were increasingly
biased as field sampling intensity decreased. In general, these results
indicate that airborne remote sensing, collected in a sampling mode, can
be used to increase the efficiency of national forest inventories.\\
\textbf{Keywords}: inventory, biomass, carbon, lidar, sampling.

\hypertarget{introduction}{%
\section{Introduction}\label{introduction}}

Reliable and credible information on forest status and change is needed
to support effective natural resource management and policy. National
forest inventory (NFI) programs are usually the primary source for this
type of information, since, ideally, they are designed to provide
accurate (unbiased) and precise estimates at regional and national
scales. For example, NFI data is used to support global forest
monitoring (FAO (2016)), national scale greenhouse gas monitoring (EPA
(2023)), and emerging markets for trading of forest carbon offset
credits (Marland et al. (2017)). The NFI is used to characterize various
attributes (e.g., biomass, area) of the population for a multitude of
domains (e.g., forest type), where estimates take the form of domain
totals (e.g., total biomass of trees in forestland), areas (e.g., total
forestland area), or per-unit-area densities (average tree biomass per
hectare of forestland). A large network of field inventory plots are the
basis for most NFI programs, and plots within homogeneous regions, or
strata, are often grouped (and weighted by stratum area) to increase the
precision of inventory estimates. Increasing interest in more accurate
estimation across spatial scales ranging from smaller regions (e.g.,
county,parcel, or management unit) (Dettmann et al. (2022)) to vast,
highly remote regions (tropical rainforests, high-latitude boreal
forests), as well as other societal drivers such the need to verify
forest carbon offset credits in carbon markets, has motivated the
international forest monitoring community to more effectively leverage
remote sensing in inventory sampling designs, including the use of
high-resolution airborne remote sensing as an intermediate level of
sampling in multi-stage sampling designs. In recent years, more advanced
approaches to estimation have been proposed that can more fully leverage
more sophisticated models and the rich information from both satellite
and airborne remote sensing in the context of NFI programs (Ståhl et al.
(2016); Lister et al. (2020); Kangas et al. (2018); McRoberts, Andersen,
and Næsset (2014); Westfall et al. (2022)). For example, both
model-assisted (Breidt and Opsomer (2017); McConville et al. (2017);
Magnussen, Nord-Larsen, and Riis-Nielsen (2018)) and model-based
techniques (McRoberts (2010)) have been developed to leverage
wall-to-wall remote sensing data in forest inventory.

The use of high-resolution, airborne lidar strip sampling to support
forest inventories has been the focus of several recent studies in
Europe and North America where a variety of estimation approaches have
been developed and evaluated, including model-assisted (Gregoire et al.
(2011); Strunk et al. (2014); Ringvall et al. (2016)), model-based
(Saarela et al. (2016)) and hybrid estimation (Ståhl et al. (2011);
Ståhl et al. (2016)). Gobakken et al. (2012) compared several
model-dependent and model-assisted estimators in a two-stage design
using lidar sampling in the Hedmark region of Norway. Ringvall et al.
(2016) developed a model-assisted estimator for two-stage lidar sample
surveys that addressed several significant limitations of lidar sampling
in the forest inventory context, including the use of a ratio estimator
to account for varying strip lengths, and poststratification to further
improve precision. This estimator has since been applied to repeat
collections of lidar sampling data to support monitoring of forest
change in Norway (Strîmbu et al. (2017)). In the context of the US NFI
in interior Alaska, model-assisted (Andersen, Barrett, et al. (2009)),
hierarchical Bayesian coregionalization model-based (Babcock et al.
(2018)) and hybrid (Ene et al. (2018)) estimation approaches have been
proposed as a means to combine information from NFI plots, a lidar strip
sample, and wall-to-wall layers.

While these more complex sampling designs can often increase the
efficiency of the inventory (higher precision at specified cost, or
level of sampling effort), they often require the development of new
estimators for inventory attributes, as well as new variance estimators
to quantify uncertainty (i.e.~variance) of the estimators. In
particular, while approaches to classification (Andersen (2009); Shoot
et al. (2021)) and estimation (McConville et al. (2017)) of domains in
the NFI context have been recently investigated, no previous studies
have developed ratio estimators for densities in a two-stage survey
design framework. Assessing the statistical properties of these new
estimators requires a simulation framework, where the true properties of
the population are fixed and known and the sampling distributions of the
estimators can be developed over many simulated samples. Assessment of
the estimators via simulation is necessary to establish the reliability
and defensibility of the estimators, especially in the context of a
national forest inventory.

The objectives of this study are to present a set of model-assisted
estimators for all types of NFI estimates required to report current
status (domain totals, areas, and densities) that can incorporate three
levels of information available in a poststratified, two-stage sampling
design: (1) a sparse sample of field inventory plots, (2) a dense strip
sample of airborne lidar, and (3) a wall-to-wall stratification layer.
This comprehensive set of estimators includes those presented in
previous studies (Ringvall et al. (2016)), as well as new
\emph{ratio-of-ratios} estimators for estimating densities.

\hypertarget{methods}{%
\section{Methods}\label{methods}}

\hypertarget{case-study-implementation-of-the-national-forest-inventory-in-interior-alaska-usa}{%
\subsection{Case study: Implementation of the National Forest Inventory
in interior Alaska,
USA}\label{case-study-implementation-of-the-national-forest-inventory-in-interior-alaska-usa}}

We focus on interior Alaska as an example of a region where an
alternative sampling design, and corresponding estimators, are needed to
support the NFI program. The US Forest Inventory and Analysis (FIA)
program -- the NFI for the United States -- is mandated by the US
Congress to assess current status and trends for all forest lands of the
United States (USDA Forest Service (2016)). The standard sampling
framework for FIA in the conterminous United States, coastal Alaska, and
Pacific islands is a systematic, unaligned grid of field plots
established with a sampling intensity of 1 plot per 2,428 ha. Interior
Alaska is a vast region of high-latitude boreal forests (approx. 46
million hectares of forestland) with virtually no transportation
infrastructure (virtually all plots are accessed via helicopter), and
for this reason, FIA has implemented a modified sampling design in
interior Alaska using a reduced sampling intensity for field plots (1
plot per 12,140 ha - 1/5\textsuperscript{th} the standard intensity --
collected on a systematic grid), supplemented with high-resolution
airborne imagery acquired by NASA Goddard's Lidar, Hyperspectral, and
Thermal (G-LiHT) Airborne Imager (Cook et al. (2013)) in a strip
sampling mode (Cahoon and Baer (2022)) covering every FIA plot. The FIA
field inventory of interior Alaska was initiated within the Tanana
inventory unit (13.5 million ha) and carried out over a five-year period
(2014-2018), and the G-LiHT collection was carried out in 2014 (see
Figure 1).

\begin{figure}[!htbp]
\centering
\includegraphics[width=16 cm]{"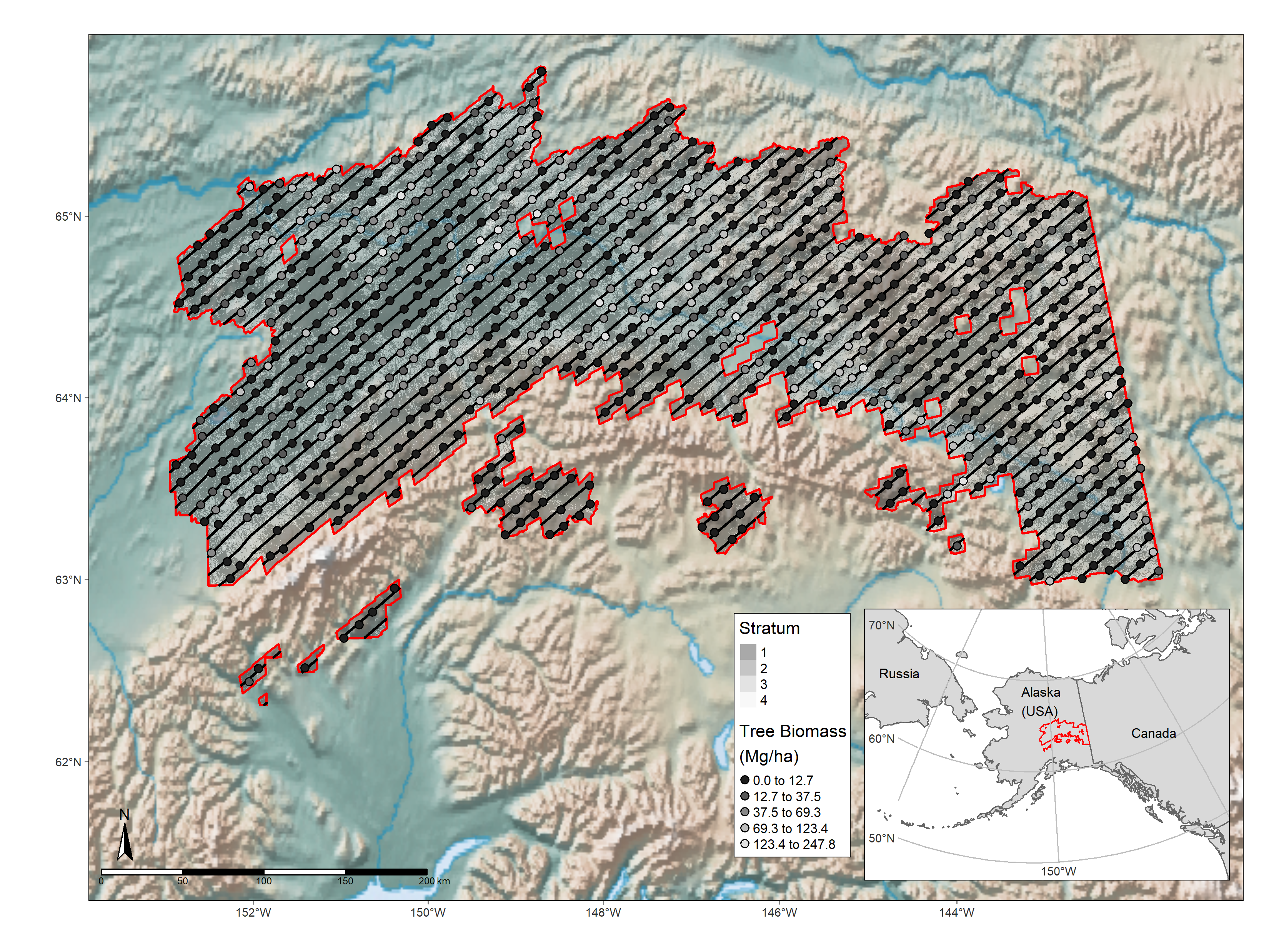"}
\caption{Tanana inventory unit, interior Alaska, USA. Black lines indicate location of airborne G-LiHT flight lines. Dots indicate (approximate) location of FIA field plots (colored by biomass). Underlying raster indicates stratification, red outline indicates the domain sampled by G-LiHT.}
\end{figure}

\hypertarget{standard-fia-sampling-design-and-estimation}{%
\subsection{Standard FIA sampling design and
estimation}\label{standard-fia-sampling-design-and-estimation}}

The sample-based estimators of population totals, areas, and densities
(and corresponding variance estimators) used in standard FIA reporting
assume that the plots are collected as a simple random sample (SRS), and
the precision of these estimators is further improved through
post-stratification, as documented in Bechtold and Patterson (2005). The
poststrata used in FIA are generally developed based on available land
cover maps, as well as ownership and other GIS layers.

\hypertarget{poststratified-estimation-of-a-domain-total}{%
\subsubsection{Poststratified estimation of a domain
total}\label{poststratified-estimation-of-a-domain-total}}

Estimates of population totals, such as tree biomass, or carbon, are
usually primary attributes of interest in a national forest inventory.
In the standard estimation framework of FIA, these values are provided
by the following estimators (including variance):

\begin{equation}
{\overline{t}}_{h}=\frac{\sum_{i = 1}^{n_{h}}y_{hi}}{n_{h}}
\end{equation}

\begin{equation}
{\widehat{t}}_{SRS,PS}=A_{T}\sum_{h = 1}^{H}W_{h} {\overline{t}}_{h}
\end{equation}

where \(y_{hi}\) is the plot-level value (per ha) for an inventory
attribute of interest (e.g., tree biomass, carbon, volume, etc.) on plot
\(i\) in stratum \(h\), \(n_{h}\) is the number of plots in stratum
\(h\), (\(N_{h}\) is the size of the stratum in the population),
\(W_{h}=N_{h}/N\) is the weight for stratum \(h\), and \(A_{T}\) is the
total area of the population (ha).

The variance estimator is:

\begin{equation}
\widehat{V}\left( {\widehat{t}}_{SRS,PS} \right)=A_{T}^{2}\left\lbrack \sum_{h = 1}^{H}W_{h}n_{h}\widehat{V}\left( {\overline{t}}_{h} \right)+\sum_{h = 1}^{H}\left(1 - W_{h} \right)\frac{n_{h}}{n}\widehat{V}\left( {\overline{t}}_{h} \right) \right\rbrack
\end{equation}

\begin{equation}
\widehat{V}\left( {\overline{t}}_{h} \right)=\frac{\sum_{i = 1}^{n_h}y_{hi}^{2}-n_h{{\overline{t}}_{h}^{2}}}{n_h(n_h- 1)}
\end{equation}

\hypertarget{poststratified-estimation-of-domain-area}{%
\subsubsection{Poststratified estimation of domain
area}\label{poststratified-estimation-of-domain-area}}

Estimation of domain area (e.g., total forestland area in hectares,
etc.) is another primary output of a national forest inventory, and are
provided by the following estimators (including variance):

\begin{equation}
{\overline{a}}_{h}=\frac{\sum_{i = 1}^{n_{h}}a_{hi}}{n_{h}}
\end{equation}

\begin{equation}
{\widehat{A}}_{SRS,PS}=A_{T}\sum_{h = 1}^{H}W_{h} {\overline{a}}_{h}
\end{equation}

where \(a_{hi}\) is the proportion of plot \(i\), in stratum \(h\),
within a given domain (e.g., forestland, forest type, etc.).

The variance estimator is:

\begin{equation}
\widehat{V}\left( {\widehat{A}}_{SRS,PS} \right)=A_{T}^{2}\left\lbrack \sum_{h = 1}^{H}W_{h}n_{h}\widehat{V}\left( {\overline{a}}_{h} \right)+\sum_{h = 1}^{H}\left(1 - W_{h} \right)\frac{n_{h}}{n}\widehat{V}\left( {\overline{a}}_{h} \right) \right\rbrack
\end{equation}

\begin{equation}
\widehat{V}\left( {\overline{a}}_{h} \right)=\frac{\sum_{i = 1}^{n_{h}}a_{hi}^{2}-n_{h}{\overline{a}}_{h}^{2}}{n_h(n_h-1)}
\end{equation}

\hypertarget{poststratified-estimation-of-a-domain-density}{%
\subsubsection{Poststratified estimation of a domain
density}\label{poststratified-estimation-of-a-domain-density}}

The final category of estimates required in a NFI include densities for
various inventory attributes on a per-unit-area basis. In the case where
a density is estimated over the entire area of the population, the
estimator is obtained by simply dividing the estimator of the total
\(\widehat{t}_{SRS,PS}\) by the known area of the population \(A_T\),
and the corresponding variance estimator is obtained by dividing the
variance of the total by \(A_T^{2}\). However, often a density is
estimated for a domain (e.g., tree carbon per forested hectare, where
the domain is forestland). Because both the numerator and denominator of
these estimators are themselves estimators, they are considered
\(ratio\) estimators. The poststratified estimator of domain density
used in FIA is given by:

\begin{equation}
{\widehat{D}}_{SRS,PS}= \frac{\sum_{h = 1}^{H}W_{h} {\overline{t}}_{h}}{\sum_{h = 1}^{H}W_{h} {\overline{a}}_{h}}
\end{equation}

with variance estimator:

\begin{equation}
\widehat{V}\left( {\widehat{D}}_{SRS,PS}\right) = \frac{1}{{\widehat{A}}_{SRS,PS}^{2}}\left\lbrack \widehat{V}\left( {\widehat{t}}_{SRS,PS}) + {\widehat{D}}_{SRS,PS}^{2}{\widehat{V}(\widehat{A}}_{SRS,PS}) \right) - 2{\widehat{D}}_{SRS,PS}\ Cov\left( {\widehat{t}}_{SRS,PS},{\widehat{A}}_{SRS,PS} \right) \right\rbrack
\end{equation}

where

\begin{equation}
Cov\left( {\widehat{t}}_{SRS,PS},{\widehat{A}}_{SRS,PS} \right) = \frac{A_{T}^{2}}{n}
\left\lbrack \sum_{h = 1}^{H}W_{h}n_{h}Cov\left( {\overline{t}}_{h}, {\overline{a}}_{h}\right) + \sum_{h = 1}^{H}\left(1-W_{h}\right)\frac{n_{h}}{n}Cov\left( {\overline{t}}_{h}, {\overline{a}}_{h}\right) \right\rbrack
\end{equation}

\begin{equation}
Cov\left( {\overline{t}}_{h},{\overline{a}}_{h} \right) = \frac{\sum_{i = 1}^{n_{h}}y_{hi}a_{hi}-n_{h}{\overline{t}}_{h} {\overline{a}}_{h}}{n_{h}\left(n_{h}-1 \right)}
\end{equation}

\hypertarget{ratio-estimators-for-model-assisted-estimation-in-two-stage-sampling-designs}{%
\subsection{Ratio estimators for model-assisted estimation in two-stage
sampling
designs}\label{ratio-estimators-for-model-assisted-estimation-in-two-stage-sampling-designs}}

\hypertarget{sampling-design-and-notation}{%
\subsubsection{Sampling design and
notation}\label{sampling-design-and-notation}}

The addition of another level of sampling, e.g., provided by
high-resolution, airborne data, collected as a strip sample, requires a
different set of domain-level estimators. The notation and formulas used
to describe these estimators mainly follow the notation and formulas
used by Ringvall et al. (2016). The study area is partitioned into
\emph{M} non-overlapping strips, denoted as \emph{PSU}s. These strips
have unequal lengths and thus unequal areas, and the set
\emph{U}\textsubscript{1} is the population of \emph{PSU}s. A
first-stage sample, \emph{S}\textsubscript{1}, of \emph{PSU}s is
selected. Each \emph{PSU} is partitioned into \emph{N\textsubscript{i}}
grid-cells (\emph{SSU}s) with a size roughly corresponding to the size
of the field plots. In the second stage, we randomly select a sample
\emph{S\textsubscript{i}} of \emph{SSU}s from each \emph{PSU}.
Subsampling within selected \emph{PSU}s are invariant of which
\emph{PSU}s are selected in the first stage. The total number of
grid-cells in the study area is \(N = \sum_{i \in U_{1}}^{}N_{i}\).

In inventory programs where the plot is actually a cluster of several
subplots (as is the case with many NFIs, including FIA), a slight
adjustment to the sampling framework can be used to maintain consistency
in the spatial support (i.e.~plot footprint) for sampling units at the
different sampling stages. To this end, in this study \emph{N} is the
number of points on the 200 meter hexagonal grid across the study area,
\(N_{i}\) is the points on this grid along the \(i^{th}\) \emph{PSU}.
The per-ha average is estimated for each attribute in \emph{N} and then
scaled up to the level of the population by multiplying by the
population area, \(A_{T}\). However, for the purposes of clarity and
consistency, we maintain the more general notation used in Ringvall et
al. (2016) in the following sections (it should be noted that the above
adjustment does not change the results).

Several other details regarding the sampling design should be noted.
First, in practice lidar strips are often allocated systematically,
traversing the entire area of study, but we assume them to be randomly
allocated from the point of view of the variance formulas. Likewise,
field plots are often distributed systematically (i.e.~at a regular
spacing) within lidar strips, we assume them to be randomly allocated
within strips, at a specified sampling intensity, from the point of view
of the variance formulas. As is the case with many standard NFI designs,
the entire study area is divided into strata, and this map information
is incorporated in the estimators through poststratification. It should
also be noted that the lidar strip sampling is carried out independent
of the stratification. In addition, the strata normally do not coincide
with the domains of interest, since areas belonging to the domain
potentially appear scattered across several strata and they cannot be
perfectly identified on maps, but only in the field on the subplots.
Prediction models for the variable of interest (e.g., biomass per ha),
as well as the domain (e.g., forestland), are developed from the
collected field data.

\hypertarget{a-ratio-estimator-of-the-domain-total}{%
\subsubsection{A ratio estimator of the domain
total}\label{a-ratio-estimator-of-the-domain-total}}

With simple random sampling without replacement in both stages, a ratio
(\emph{R}) estimator of the variable of interest within a given domain
(cf. Ringvall et al. (2016), eq. 5) is

\begin{equation}
\widehat{t}_{R} = N\frac{\sum_{i = 1}^{m}{\ {\widehat{t}}_{ri}}}{\sum_{i = 1}^{m}N_{i}}
\end{equation}

where \emph{m} is the number of selected \emph{PSU}s and
\({\widehat{t}}_{ri}\) is the model-assisted estimator of total biomass
in a given domain (where the \emph{r} subscript denotes a regression
estimator) in the \(i^{th}\) \emph{PSU} given as

\begin{equation}
\widehat{t}_{ri} = \sum_{k = 1}^{N_{i}}{{\widehat{q}}_{k}{\widehat{y}}_{k}} + \frac{N_{i}}{n_{i}}\sum_{k = 1}^{n_{i}}e_{k}
\end{equation}

where \emph{n\textsubscript{i}} is the number of selected \emph{SSU}s
within \emph{PSU} \emph{i}, \({\widehat{y}}_{k}\) is a predicted value
of the domain-level attribute of interest based on a regression (or
similar) model within a grid cell/plot, \(\widehat{q}_{k}\) is an
indicator variable (e.g., based on logistic regression) predicting
whether or not the grid cell/plot belongs to the domain (or a prediction
of the proportion in a domain, e.g., Van den Boogaart and
Tolosana-Delgado (2013)), and \(e_{k}\) is a deviation between the
measured and predicted plot-level value for the variable of interest in
the domain. In cases where the domain membership is incorporated into
the regression model predicting the variable of interest (e.g., biomass
is always zero on ``nonforest'' plots), the \(\widehat{q}_{k}\) term can
be omitted. It should be noted that predictions of the target variable
and domain membership are made at the scale of the entire plot. If
predictions are made at subplot scale in the case of cluster plots, it
is assumed that these predictions, and associated error terms, are
averaged and aggregated to the plot scale.

The variance estimator of \({\widehat{t}}_{R}\) is

\begin{equation}
\widehat{V}\left( {\widehat{t}}_{R} \right) = \frac{N^{2}}{{\widehat{N}}^{2}}\left\lbrack M^{2}\left( \frac{1}{m} - \frac{1}{M} \right)s_{r}^{2} + \frac{M}{m}\sum_{S_{1}}^{}N_{i}^{2}\left( \frac{1}{n_{i}} - \frac{1}{N_{i}} \right)s_{e}^{2} \right\rbrack
\end{equation}

where
\(s_{r}^{2} = \frac{1}{m - 1}\sum_{i = 1}^{m}{({\widehat{t}}_{ri} - \widehat{R}N_{i})}^{2}\)
with
\(\widehat{R} = \frac{\sum_{i = 1}^{m}{\ {\widehat{t}}_{ri}}}{\sum_{i = 1}^{m}N_{i}}\)
and
\(s_{e}^{2} = \frac{1}{n_{i} - 1}\sum_{k = 1}^{n_{i}}\left( e_{k} - {\overline{e}}_{i} \right)^{2}\)
with \({\overline{e}}_{i} = \frac{1}{n_{i}}\sum_{k = 1}^{n_{i}}e_{k}\).

\hypertarget{a-ratio-estimator-of-the-domain-total-in-poststratified-estimation}{%
\subsubsection{A ratio estimator of the domain total in poststratified
estimation}\label{a-ratio-estimator-of-the-domain-total-in-poststratified-estimation}}

When \emph{m} \emph{PSU}s are selected with simple random sampling
without replacement in the first stage and \emph{n\textsubscript{i}}
units are selected with simple random sampling with replacement in the
second stage (as before), an estimator of a stratum total is

\begin{equation}
\widehat{t}_{Rh} = N_{h}\frac{\sum_{i = 1}^{m}{\widehat{t}}_{rhi}}{\sum_{i = 1}^{m}N_{hi}}
\end{equation}

with

\begin{equation}
\widehat{t}_{rhi} = \sum_{k = 1}^{N_{hi}}{{\widehat{q}}_{hk}{\widehat{y}}_{hk}} + \frac{N_{hi}}{n_{hi}}\sum_{k = 1}^{n_{hi}}e_{hk}
\end{equation}

where the notation follows from the previously used notation, but refers
to a specific stratum \emph{h}. Note that \(e_{hk}\) involves either (1)
prediction of domain membership for subplots and averaging across
subplots, or (2) prediction of domain membership for the entire plot. In
addition, all subplots are assumed to belong to the same stratum.

The poststratified estimator of the domain total is:

\begin{equation}
{\widehat{t}}_{R,PS} = \sum_{h = 1}^{H}{\widehat{t}}_{Rh}
\end{equation}

The variance estimator becomes (from Ringvall et al. (2016))

\begin{equation}
\widehat{V}({\widehat{t}}_{Rh}) = \left( \frac{N_{h}}{{\widehat{N}}_{h}} \right)^{2}\left( M^{2}\left( \frac{1}{m} - \frac{1}{M} \right)s_{rh}^{2} + \frac{M}{m}\sum_{S_{1}}^{}{{\widehat{V}(\widehat{t}}_{rhi})} \right)
\end{equation}

where
\(s_{rh}^{2} = \frac{1}{m - 1}\sum_{i = 1}^{m}\left( {\widehat{t}}_{rhi} - {\widehat{R}}_{h}N_{hi} \right)^{2}\)
with
\({\widehat{R}}_{h} = \frac{\sum_{S_{I}}^{}{\widehat{t}}_{rhi}}{\sum_{S_{I}}^{}N_{hi}}\).
The within PSU variance \(V({\widehat{t}}_{rhi})\) is estimated as

\begin{equation}
\widehat{V}(\widehat{t}_{rhi}) = N_{hi}^{2}\left( \frac{1}{n_{hi}} - \frac{1}{N_{hi}} \right)s_{{\widehat{e}}_{hi}}^{2}
\end{equation}

with
\(s_{e,h}^{2} = \frac{1}{n_{hi} - 1}\sum_{k = 1}^{n_{hi}}{({e_{hk} - {\overline{e}}_{hi})}^{2}}\)
and
\({\overline{e}}_{hi} = \frac{1}{n_{hi}}\sum_{k = 1}^{n_{hi}}e_{hk}\).

Finally, the poststratified estimator of the domain total is
\({\widehat{t}}_{R,PS} = \sum_{h = 1}^{H}{N_{h}\frac{\sum_{i = 1}^{m}{\widehat{t}}_{rhi}}{\sum_{i = 1}^{m}N_{hi}}}\).
The variance estimator is

\begin{equation}
\widehat{V}\left( {\widehat{t}}_{R,PS} \right) = \sum_{h = 1}^{H}{{\widehat{V}(\widehat{t}}_{Rh}) + \sum_{h = 1}^{H}{\sum_{g \neq h}^{}{\frac{N_{h}}{{\widehat{N}}_{h}}\frac{N_{g}}{{\widehat{N}}_{g}}M^{2}\left( \frac{1}{m} - \frac{1}{M} \right)\frac{\sum_{i = 1}^{m}\left( \left( {\ \widehat{t}}_{rhi} - {\widehat{R}}_{h}N_{hi} \right)\left( {\ \widehat{t}}_{rgi} - {\widehat{R}}_{g}N_{gi} \right) \right)}{m - 1}}}}
\end{equation}

with \({\widehat{V}(\widehat{t}}_{Rh})\) given by Eq. 19. In this case
the computations can be limited to those strata where the domain has a
potential to occur. Note that because in our case sampling in not
conducted independently between strata, the variance estimator (Eq. 21)
must account for covariances among strata (i.e.~the last term on the
right hand side).

\hypertarget{a-ratio-estimator-of-the-domain-area}{%
\subsubsection{A ratio estimator of the domain
area}\label{a-ratio-estimator-of-the-domain-area}}

So far, the population total within the domain has been addressed.
Following the previous notation and estimation principles, a domain area
estimator (without poststratification) would be

\begin{equation}
\widehat{A}_{R} = N\frac{\sum_{i = 1}^{m}{\ {\widehat{a}}_{ri}}}{\sum_{i = 1}^{m}N_{i}}
\end{equation}

where \({\widehat{a}}_{ri}\) is the model-assisted estimator of total
domain area in the \(i^{th}\) \emph{PSU} given as

\begin{equation}
\widehat{a}_{ri} = \sum_{k = 1}^{N_{i}}{{\widehat{q}}_{k}\ a\_ cell} + \frac{N_{i}}{n_{i}}\sum_{k = 1}^{n_{i}}e_{k}
\end{equation}

where \(a\_ cell\) is the area of a grid-cell. The other notation is the
same as previously, and notably \(e_{k}\) is a deviation between a
measured and predicted plot/grid-cell level value for the proportion of
the plot within the domain. Like for the target variable of interest,
\(e_{k}\) should be the aggregated and averaged difference across
subplots, recalculated to correspond to the size of a grid-cell.

The variance estimator of \({\widehat{A}}_{R}\) is

\begin{equation}
\widehat{V}\left( {\widehat{A}}_{R} \right) = \frac{N^{2}}{{\widehat{N}}^{2}}\left\lbrack M^{2}\left( \frac{1}{m} - \frac{1}{M} \right)s_{r}^{2} + \frac{M}{m}\sum_{S_{1}}^{}N_{i}^{2}\left( \frac{1}{n_{i}} - \frac{1}{N_{i}} \right)s_{e}^{2} \right\rbrack
\end{equation}

where
\(s_{r}^{2} = \frac{1}{m - 1}\sum_{i = 1}^{m}{({\widehat{a}}_{ri} - \widehat{R}N_{i})}^{2}\)
with
\(\widehat{R} = \frac{\sum_{i = 1}^{m}{\ {\widehat{a}}_{ri}}}{\sum_{i = 1}^{m}N_{i}}\)
and
\(s_{e}^{2} = \frac{1}{n_{i} - 1}\sum_{k = 1}^{n_{i}}\left( e_{k} - {\overline{e}}_{i} \right)^{2}\)
with \({\overline{e}}_{i} = \frac{1}{n_{i}}\sum_{k = 1}^{n_{i}}e_{k}\)..

\hypertarget{a-ratio-estimator-of-the-domain-area-in-poststratified-estimation}{%
\subsubsection{A ratio estimator of the domain area in poststratified
estimation}\label{a-ratio-estimator-of-the-domain-area-in-poststratified-estimation}}

So far, the population total within the domain has been addressed.
Following the previous notation and estimation principles, a domain area
estimator (with poststratification) would be

\begin{equation}
\widehat{A}_{Rh} = N_{h}\frac{\sum_{i = 1}^{m}{\widehat{a}}_{rhi}}{\sum_{i = 1}^{m}N_{ih}}
\end{equation}

where \({\widehat{a}}_{rhi}\) is the model-assisted estimator of total
domain area, in stratum \emph{h}, in the \(i^{th}\) \emph{PSU} given as

\begin{equation}
\widehat{a}_{rhi} = \sum_{k = 1}^{N_{hi}}{{\widehat{q}}_{hk}\ a\_ cell} + \frac{N_{hi}}{n_{hi}}\sum_{k = 1}^{n_{hi}}e_{hk}
\end{equation}

where \(a\_ cell\) is the area of a grid-cell.

The poststratified estimator of the total population area is:

\begin{equation}
{\widehat{A}}_{R,PS} = \sum_{h = 1}^{H}{\widehat{A}}_{Rh}
\end{equation}

with variance estimator

\begin{equation}
\widehat{V}\left( {\widehat{A}}_{Rh} \right) = \left( \frac{N_{h}}{{\widehat{N}}_{h}} \right)^{2}\left( M^{2}\left( \frac{1}{m} - \frac{1}{M} \right)s_{rh}^{2} + \frac{M}{m}\sum_{S_{1}}^{}{{\widehat{V}(\widehat{a}}_{rhi})} \right)
\end{equation}

where
\(s_{rh}^{2} = \frac{1}{m - 1}\sum_{i = 1}^{m}\left( {\widehat{a}}_{rhi} - {\widehat{R}}_{a,h}N_{hi} \right)^{2}\)
and
\({\widehat{R}}_{a,h} = \frac{\sum_{S_{1}}^{}{\widehat{a}}_{rhi}}{\sum_{S_{1}}^{}N_{hi}}\).
\(V({\widehat{t}}_{rhi})\)is estimated as

\begin{equation}
\widehat{V}(\widehat{a}_{rhi}) = N_{hi}^{2}\left( \frac{1}{n_{hi}} - \frac{1}{{\widehat{N}}_{hi}} \right)s_{e,h}^{2}
\end{equation}

with
\(s_{e,h}^{2} = \frac{1}{n_{hi} - 1}\sum_{k = 1}^{n_{hi}}{({e_{hk} - {\overline{e}}_{hi})}^{2}}\)
and
\({\overline{e}}_{hi} = \frac{1}{n_{hi}}\sum_{k = 1}^{n_{hi}}e_{hk}\).

A variance estimator for the total domain area is given by:

\begin{equation}
\widehat{V}\left( {\widehat{A}}_{R,PS} \right) = \sum_{h = 1}^{H}{{\widehat{V}(\widehat{A}}_{Rh}) + \sum_{h = 1}^{H}{\sum_{g \neq h}^{}{\frac{N_{h}}{{\widehat{N}}_{h}}\frac{N_{g}}{{\widehat{N}}_{g}}M^{2}\left( \frac{1}{m} - \frac{1}{M} \right)\frac{\sum_{i = 1}^{m}\left( \left( {\ \widehat{a}}_{rhi} - {\widehat{R}}_{a,h}N_{hi} \right)\left( {\ \widehat{a}}_{rgi} - {\widehat{R}}_{a,g}N_{gi} \right) \right)}{m - 1}}}}
\end{equation}

with \({\widehat{V}(\widehat{A}}_{Rh})\) given by Eq. 27.

\hypertarget{a-ratio-of-ratios-estimator-of-the-domain-density}{%
\subsubsection{A ratio-of-ratios estimator of the domain
density}\label{a-ratio-of-ratios-estimator-of-the-domain-density}}

In the context of forest inventory, we are often also interested in
domain densities, in addition to domain totals and areas. Because the
form of the estimator of domain density is a ratio estimator, where the
numerator and denominator \emph{are themselves ratio estimators}, we
term the form of this new estimator as a \emph{ratio-of-ratios}, or
\emph{RoR}, estimator.

\begin{equation}
\widehat{D}_{RoR} = \frac{{\widehat{t}}_{R}}{{\widehat{A}}_{R}} = \ \frac{N\frac{\sum_{i = 1}^{m}{\ {\widehat{t}}_{ir}}}{\sum_{i = 1}^{m}N_{i}}}{N\frac{\sum_{i = 1}^{m}{\ {\widehat{a}}_{ir}}}{\sum_{i = 1}^{m}N_{i}}}
= \frac{\sum_{i = 1}^{m}{\ {\widehat{t}}_{ir}}}{\sum_{i = 1}^{m}{\widehat{a}}_{ir}}
\end{equation}

where \({\widehat{t}}_{R}\) is the ratio domain biomass estimator and
\({\widehat{A}}_{R}\) is the ratio domain area estimator. Introducing
the new variables \(u_{i} =\)
\({\widehat{t}}_{i} - {\widehat{D}}_{RoR}{\widehat{a}}_{i}\) and
\(v_{ik} = e_{ik}^{y} - {\widehat{D}}_{RoR}{\widehat{a}}_{ik}\) the
variance estimator for \({\widehat{D}}_{RoR}\) is

\begin{equation}
\widehat{V}\left( {\widehat{D}}_{RoR} \right) = \frac{1}{\widehat{A^*}^{2}}\left\lbrack M^{2}\left( \frac{1}{m} - \frac{1}{M} \right)s_{u}^{2} + \frac{M}{m}\sum_{S_{1}}^{}N_{i}^{2}\left( \frac{1}{n_{i}} - \frac{1}{N_{i}} \right)s_{vi}^{2} \right\rbrack
\end{equation}

where
\(s_{u}^{2} = \frac{1}{m - 1}\sum_{i = 1}^{m}{({u}_{i} - \overline{u})}^{2}\),
\(s_{vi}^{2} = \frac{1}{n_{i} - 1}\sum_{k = 1}^{n_{i}}\left( {v}_{ik} - \overline{v}_{i} \right)^{2}\),
and the term \(\widehat{A^*}\) is computed as the average estimated area
from the sampled strips, times the total number of strips in the
population, \emph{M}. Details regarding the derivation of
\(\widehat{V}\left( {\widehat{D}}_{RoR} \right)\) are provided in
Appendix A.

\hypertarget{a-ratio-of-ratios-estimator-of-the-domain-density-with-poststratification}{%
\subsubsection{A ratio-of-ratios estimator of the domain density with
poststratification}\label{a-ratio-of-ratios-estimator-of-the-domain-density-with-poststratification}}

The ratio-of-ratios estimator takes on a more complex form in a
poststratified (or prestratified) setting, since in the case of
poststratification we must account for dependencies (i.e.~covariance)
between the estimators for different strata, since a single survey line
may cross over several strata.

The ratio-of-ratios estimator with poststratification is:

\begin{equation}
\widehat{D}_{RoR,PS} = \frac{{\widehat{t}}_{R,PS}}{{\widehat{A}}_{R,PS}} = \frac{\sum_{h}^{}{\widehat{t}}_{Rh}}{\sum_{h}^{}{\widehat{A}}_{Rh}} = \ \frac{\sum_{h}^{}{N_{h}\frac{\sum_{i = 1}^{m}{\ {\widehat{t}}_{rhi}}}{\sum_{i = 1}^{m}N_{hi}}}}{\sum_{h}^{}{N_{h}\frac{\sum_{i = 1}^{m}{\ {\widehat{a}}_{rhi}}}{\sum_{i = 1}^{m}N_{hi}}}}
\end{equation}

Note that summation in (33) is across all sample strips, even if a
stratum is not present in that strip. In such cases, the contribution
from the strip is zero. Alternatively, (see Ringvall et al. (2016)) an
estimator could be based on summation only across those strips that
contain the stratum, in which case the basic estimator (33) would remain
the same but the variance and variance estimators would be slightly
different. In the following, the variances are developed for the case of
summation across all strips.

Details related to the derivation of the variance estimator of
\(\widehat{D}_{RoR,PS}\) are provided in Appendix B, which results in
the following:

\begin{multline}
\widehat{V}\left( {\widehat{D}}_{RoR,PS} \right) \approx \frac{1}{{\widehat{A}}_{R,PS}^{2}}\widehat{V}\left( {\widehat{t}}_{R,PS} - {\widehat{D}}_{RoR,PS}{\widehat{A}}_{R,PS} \right) = \\ \frac{1}{{\widehat{A}}_{R,PS}^{2}}\left\lbrack \widehat{V}\left( {\widehat{t}}_{R,PS}) + {\widehat{D}}_{RoR,PS}^{2}{\widehat{V}(\widehat{A}}_{R,PS}) \right) - 2{\widehat{D}}_{RoR,PS}\ \widehat{Cov}\left( {\widehat{t}}_{R,PS},{\widehat{A}}_{R,PS} \right) \right\rbrack
\end{multline}

with

\begin{multline}
\widehat{V}\left( {\widehat{t}}_{R,PS} \right) = \sum_{h}^{}{\ \frac{N_{h}^{2}}{{\widehat{N}}_{h}^{2}}\left\lbrack M^{2}\left( \frac{1}{m} - \frac{1}{M} \right)s_{rh}^{2} + \frac{M}{m}\sum_{S1}^{}N_{hi}^{2}\left( \frac{1}{n_{hi}} - \frac{1}{N_{hi}} \right)s_{e,h}^{2} \right\rbrack} + \\ 
\sum_{h}^{}{\sum_{g \neq h}^{}{\frac{N_{h}N_{g}}{{\widehat{N}}_{h}{\widehat{N}}_{g}}M^{2}\left( \frac{1}{m} - \frac{1}{M} \right)\widehat{Cov}\left( r_{h},r_{g} \right)\ }}
\end{multline}

and

\begin{multline}
{\widehat{V}(\widehat{A}}_{R,PS}) = \sum_{h}^{}{\frac{N_{h}^{2}}{{\widehat{N}}_{h}^{2}}\left\lbrack M^{2}\left( \frac{1}{m} - \frac{1}{M} \right)s_{ra,h}^{2} + \frac{M}{m}\sum_{S1}^{}N_{hi}^{2}\left( \frac{1}{n_{hi}} - \frac{1}{N_{hi}} \right)s_{ea,h}^{2} \right\rbrack} + \\ 
\sum_{h}^{}{\sum_{g \neq h}^{}{\frac{N_{h}N_{g}}{{\widehat{N}}_{h}{\widehat{N}}_{g}}M^{2}\left( \frac{1}{m} - \frac{1}{M} \right)\widehat{Cov}\left( r_{a,h},r_{a,g} \right)\ }}
\end{multline}

and

\begin{equation}
\widehat{Cov}\left( {\widehat{t}}_{R,PS},{\widehat{A}}_{R,PS} \right) = \sum_{h}^{}{\sum_{g}^{}{\frac{N_{h}N_{g}}{{\widehat{N}}_{h}{\widehat{N}}_{g}}M^{2}\left( \frac{1}{m} - \frac{1}{M} \right)\widehat{Cov}\left( r_{h},r_{a,g} \right)}}
\end{equation}

where \(r_{h} = {\widehat{t}}_{rhi} - {\widehat{R}}_{h}N_{hi}\) and
\(r_{a,h} = {\widehat{a}}_{rhi} - {\widehat{R}}_{a,h}N_{hi}\), and the
covariance terms estimated in the usual way across the \(m\) survey
strips.

\hypertarget{evaluation}{%
\section{Evaluation}\label{evaluation}}

\hypertarget{empirical-material-and-artificial-population}{%
\subsection{Empirical material and artificial
population}\label{empirical-material-and-artificial-population}}

\hypertarget{fia-field-plot-measurements}{%
\subsubsection{FIA field plot
measurements}\label{fia-field-plot-measurements}}

The standard FIA plot is made up of four 1/60\textsuperscript{th} ha
(7.3 m radius) fixed-area, circular subplots spaced 36.6 meters apart
(for details on the FIA plot design and measurements, see Cahoon and
Baer (2022)).The coordinates of each subplot were obtained (with less
than 2 meters error) with GLONASS-enabled Trimble GeoXH mapping-grade
GNSS receivers (McGaughey et al. (2017); Andersen, Clarkin, et al.
(2009); Andersen, Strunk, and McGaughey (2022)). Aboveground dry biomass
(Mg) for each measured tree was calculated using standard FIA
procedures, and aggregated at the plot level (Mg/ha). In addition, the
proportion of the FIA plot classified as accessible forestland was
recorded as a fractional value between 0 and 1 (inclusive).

\hypertarget{g-liht-airborne-laser-scanning-sampling}{%
\subsubsection{G-LiHT airborne laser scanning
sampling}\label{g-liht-airborne-laser-scanning-sampling}}

To augment the relatively sparse sample of field plots, single, linear
swaths of high-resolution airborne remote sensing measurements, placed
approximately 9.2 km apart, oriented in a NW-SW direction and covering
every FIA plot, were acquired using G-LiHT mounted on a fixed-wing
(Piper Cherokee) platform (Cook et al. (2013)) (see Figure 1). In this
study, only the airborne lidar data from the G-LiHT instrument was
utilized. The specifications for the lidar data collection are provided
in Table 1.

\begin{table}[H]
\caption{G-LiHT lidar specifications}
\centering
%% \tablesize{} %% You can specify the fontsize here, e.g.,  \tablesize{\footnotesize}. If commented out \small will be used.
\begin{tabular}{cc}
\toprule
Instrument & Riegl VQ-480\\
Laser wavelength & 1550 nm\\
Flying height & 335 m AGL\\
Beam divergence & 0.3 mrad\\
Footprint size & 10 cm\\
Half-scanning angle & 15 degrees\\
Average pulse density & 3 pulses/m$^2$ \\
Swath width & 400 m\\
\bottomrule
\end{tabular}
\end{table}

Lidar point cloud data was processed into 1 m resolution Digital Terrain
and Canopy Height Models (DTM and CHM, respectively) for actual measured
and simulated FIA field plots (1/15\textsuperscript{th} ha each) spaced
200 m apart on G-LiHT transects, resulting in a total of 59,090 total
remote sensing plots (Figure 1). In order to allow for a direct
comparison between high-resolution lidar-based canopy heights and the
field-measured FIA inventory parameter (e.g., tree biomass), and to
avoid any bias or inefficiency due to a difference in spatial support
regions between the FIA plot footprint and the airborne lidar
measurements, in this study the lidar was extracted within a discrete
number of ``remote sensing plots'' corresponding exactly to the FIA plot
footprint (i.e.~four 1/60\textsuperscript{th} ha subplots, etc.), and
spaced at 200 meter intervals along the center of the G-LiHT swaths. The
approach resulted in 73,509 total remote sensing plots within the Tanana
G-LIHT coverage(Figure 1) , the average of the lidar CHM grid-cells
within a FIA plot footprint (1/15\textsuperscript{th} ha) was calculated
for (1) each actual field-measured FIA plot, and (2) remote sensing
plots distributed over the entire lidar coverage area.

\hypertarget{stratification-layer}{%
\subsubsection{Stratification layer}\label{stratification-layer}}

A stratification layer was generated from the National Land Cover
Database (Homer et al. (2015)) by reclassifying ``Evergreen Forest''
(Class 42) as Class 1, ``Mixed Forest'' (class 43) as class 2,
``Deciduous Forest'' (class 41) as Class 3, and all other classes as
Class 0. These strata were chosen so that stratum was an ordinal
variable and biomass has a positive correlation with stratum, a
desirable feature in developing the simulated population (described in
next section).

\hypertarget{artificial-population}{%
\subsubsection{Artificial population}\label{artificial-population}}

Simulation can be a useful approach to gain insight into the statistical
properties of various survey estimators, especially in the case of
somewhat complex, multi-level sampling designs(Ene et al. (2012);
Saarela et al. (2017)). When generating a simulated environmental
population, it is desirable to include realistic correlations between
the response variable (e.g., biomass, forest type, etc.) and the
predictor variables used in the estimator (e.g., lidar height, strata,
etc.), realistic marginal distributions for the attributes, and
realistic spatial variation across multiple scales.

\begin{figure}[H]
\centering
\includegraphics[width=15 cm]{"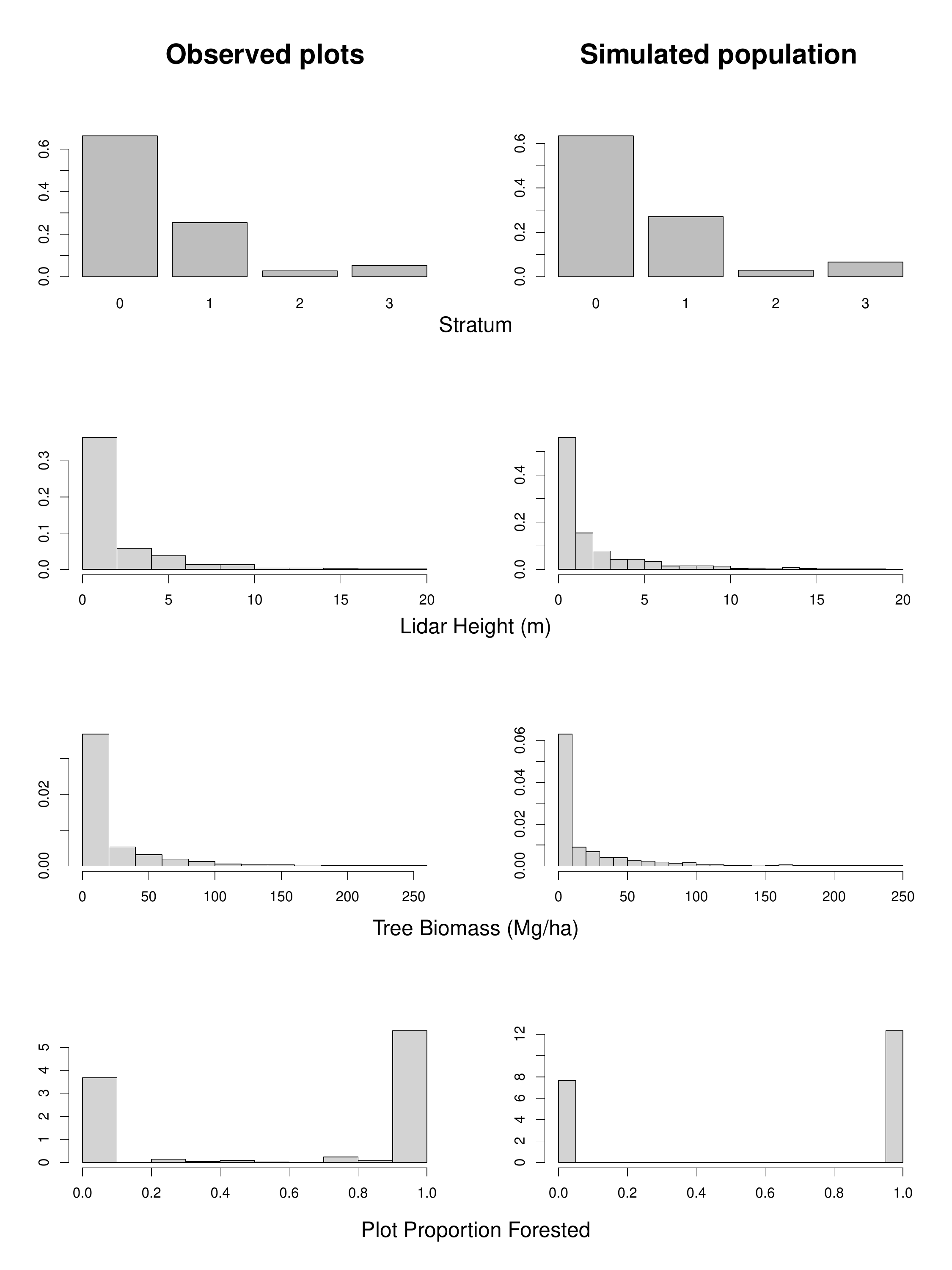"}
\caption{Marginal distributions from Observed Plots (left column) and Simulated Population (right column).}
\end{figure}

\begin{table}[H]
    \caption{Correlation Matrices for a) Observed field data, and b) Simulated population}
    \begin{subtable}{.5\linewidth}
      \centering
        \caption{}
      \begin{tabular}{c|cccc} % <-- Alignments: 1st column left, 2nd middle and 3rd right, with vertical lines in between
        {} &      Stratum & Lidar Ht. &   $y$ &  $a$\\
      \hline
      Stratum  &  1.00 & 0.64 & 0.58 & 0.41\\
      Lidar Ht.   &   0.64 & 1.00 & 0.89 & 0.42\\
      $y$     &     0.58 & 0.89 & 1.00 & 0.42\\
      $a$     &     0.41 & 0.42 & 0.42 & 1.00\\
    \end{tabular}
    \end{subtable}%
    \begin{subtable}{.5\linewidth}
      \centering
        \caption{}
      \begin{tabular}{c|cccc} % <-- Alignments: 1st column left, 2nd middle and 3rd right, with vertical lines in between
         {} &      Stratum & Lidar Ht. &   $y$ &  $a$\\
      \hline
      Stratum  &   1.00 &   0.53  & 0.60  & 0.28\\
      Lidar Ht.   &    0.53 &   1.00  & 0.84  & 0.42\\
      $y$     &    0.60 &   0.84  & 1.00  & 0.37\\
      $a$     &    0.28 &   0.42  & 0.37  & 1.00\\
    \end{tabular}
    \end{subtable} 
\end{table}

\begin{figure}[H]
\centering
\includegraphics[width=16 cm]{"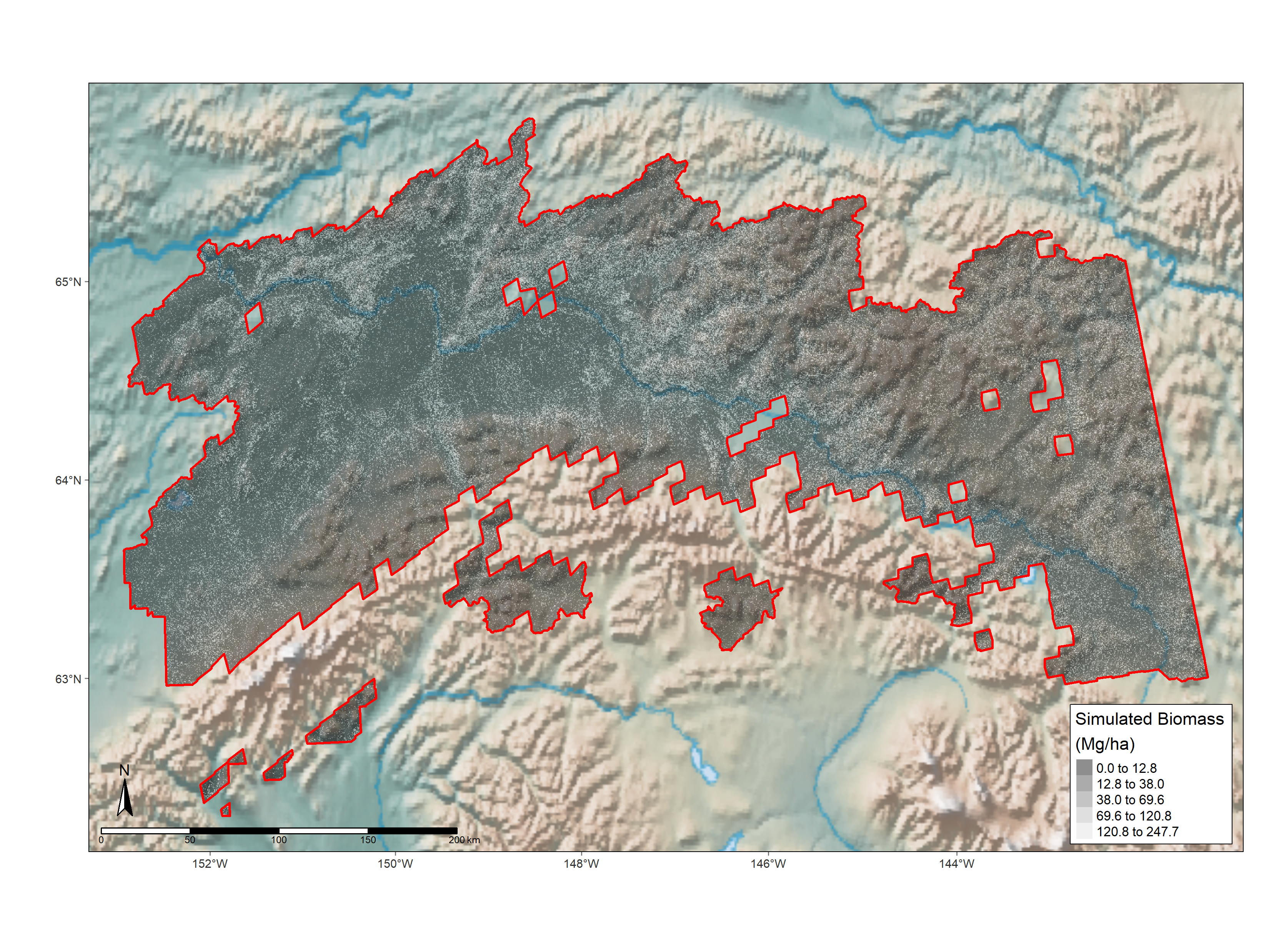"}
\caption{Spatial distribution of simulated population within Tanana study area}
\end{figure}

\begin{figure}[H]
\centering
\begin{subfigure}{0.6\textwidth}
    \includegraphics[width=\textwidth]{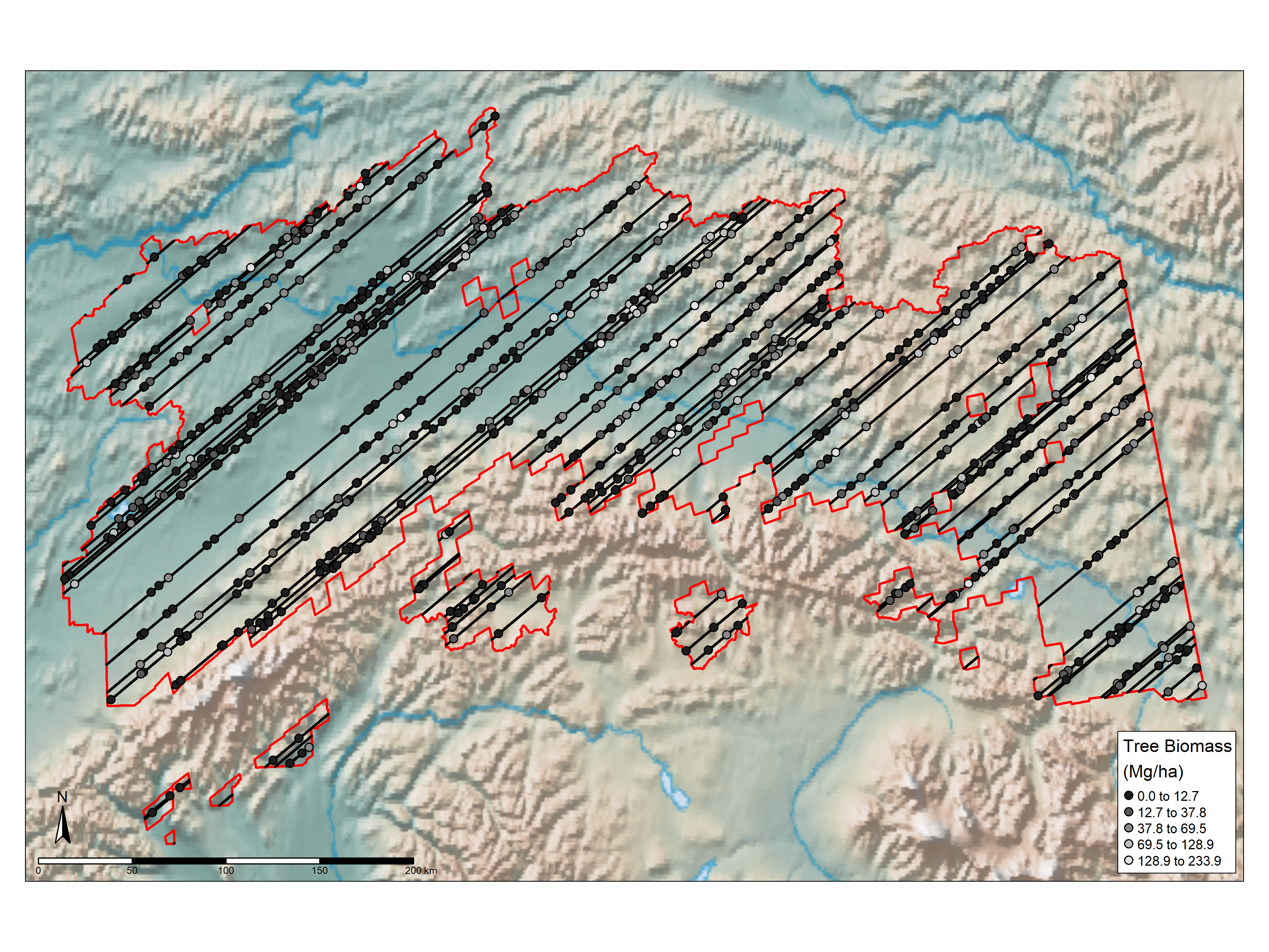}
    \caption{Field plot sampling intensity: 0.015}
    \label{fig:first}
\end{subfigure}
\begin{subfigure}{0.6\textwidth}
    \includegraphics[width=\textwidth]{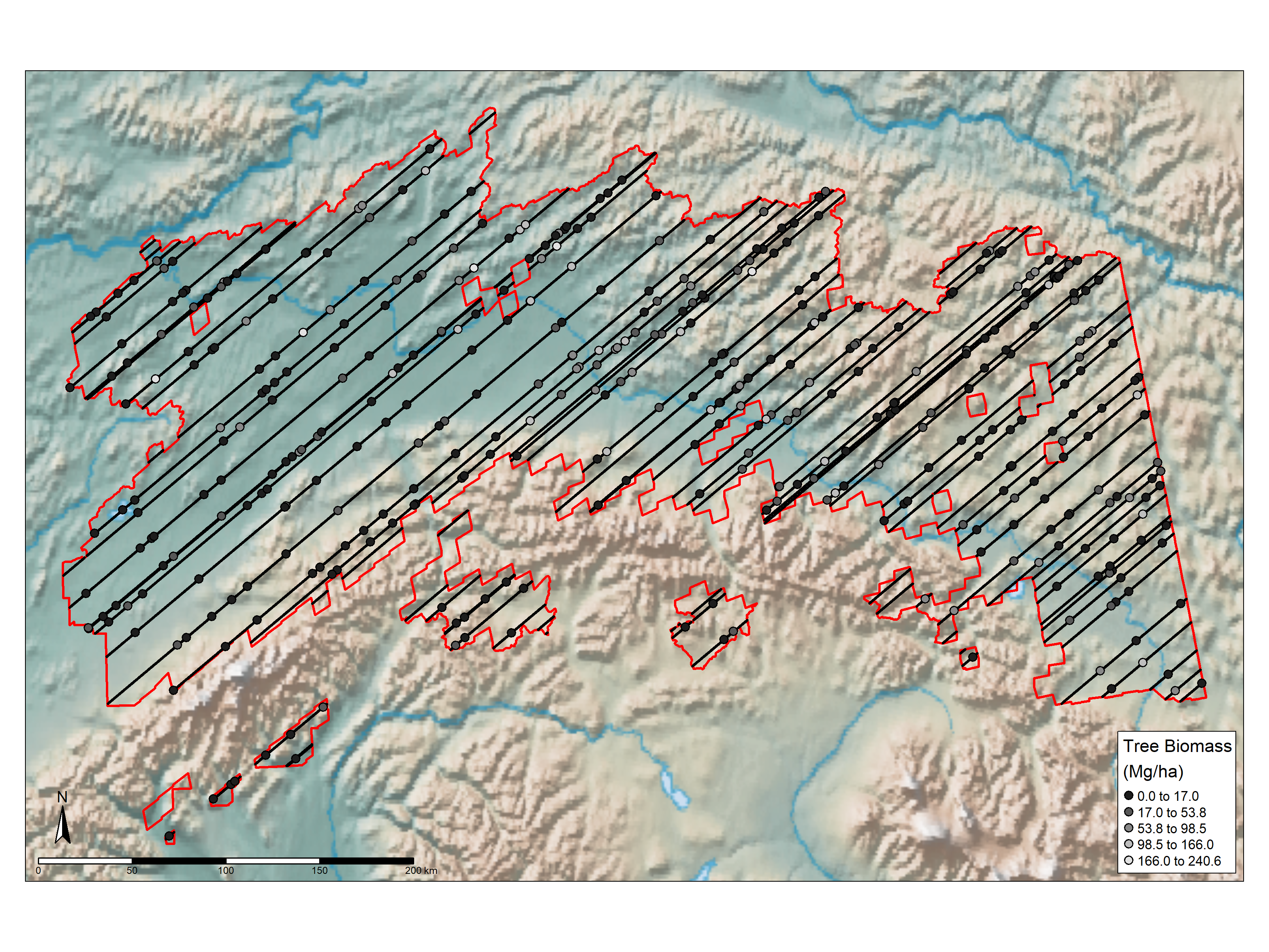}
    \caption{Field plot sampling intensity: 0.0075}
    \label{fig:second}
\end{subfigure}
\begin{subfigure}{0.6\textwidth}
    \includegraphics[width=\textwidth]{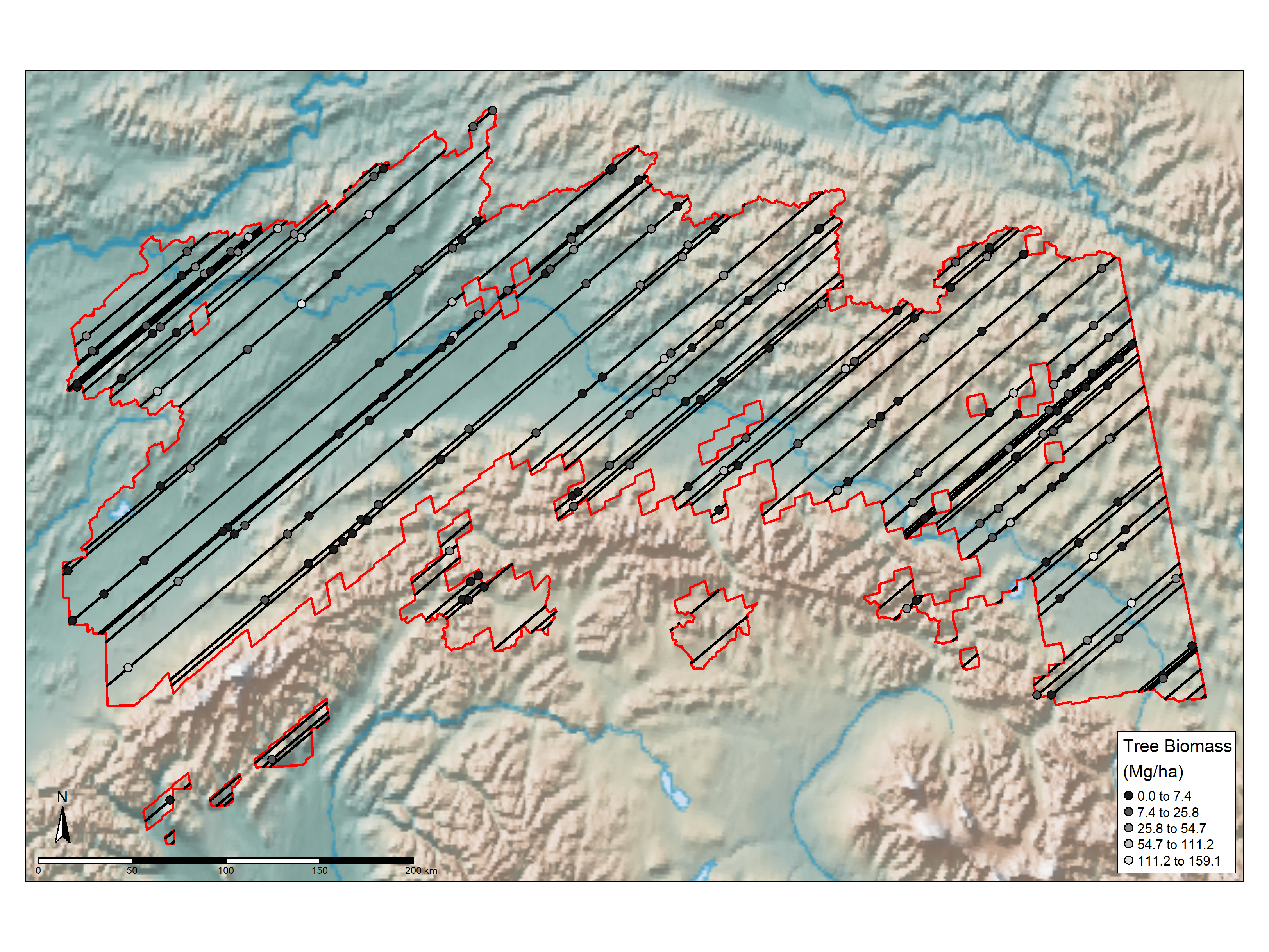}
    \caption{Field plot sampling intensity: 0.00375}
    \label{fig:third}
\end{subfigure}

\caption{Simulated samples at different sampling intensities with simple random sampling. Black lines indicate random sampled of airborne lidar flight lines. Dots indicate simulated field plots, color-coded by biomass}
\label{fig:figures}
\end{figure}

\begin{figure}[H]
\centering
\begin{subfigure}{0.6\textwidth}
    \includegraphics[width=\textwidth]{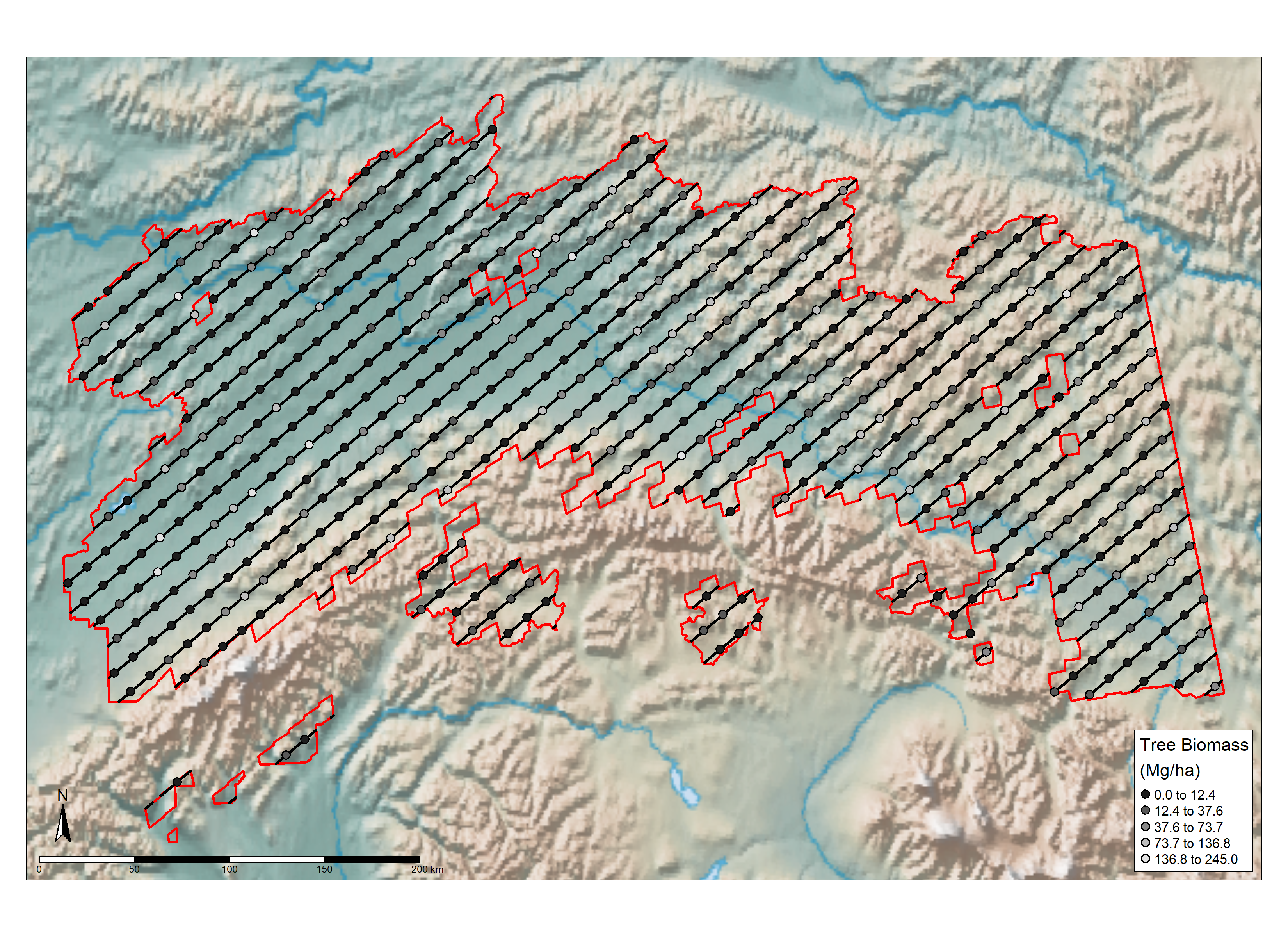}
    \caption{Field plot sampling intensity: 0.015}
    \label{fig:first}
\end{subfigure}
\begin{subfigure}{0.6\textwidth}
    \includegraphics[width=\textwidth]{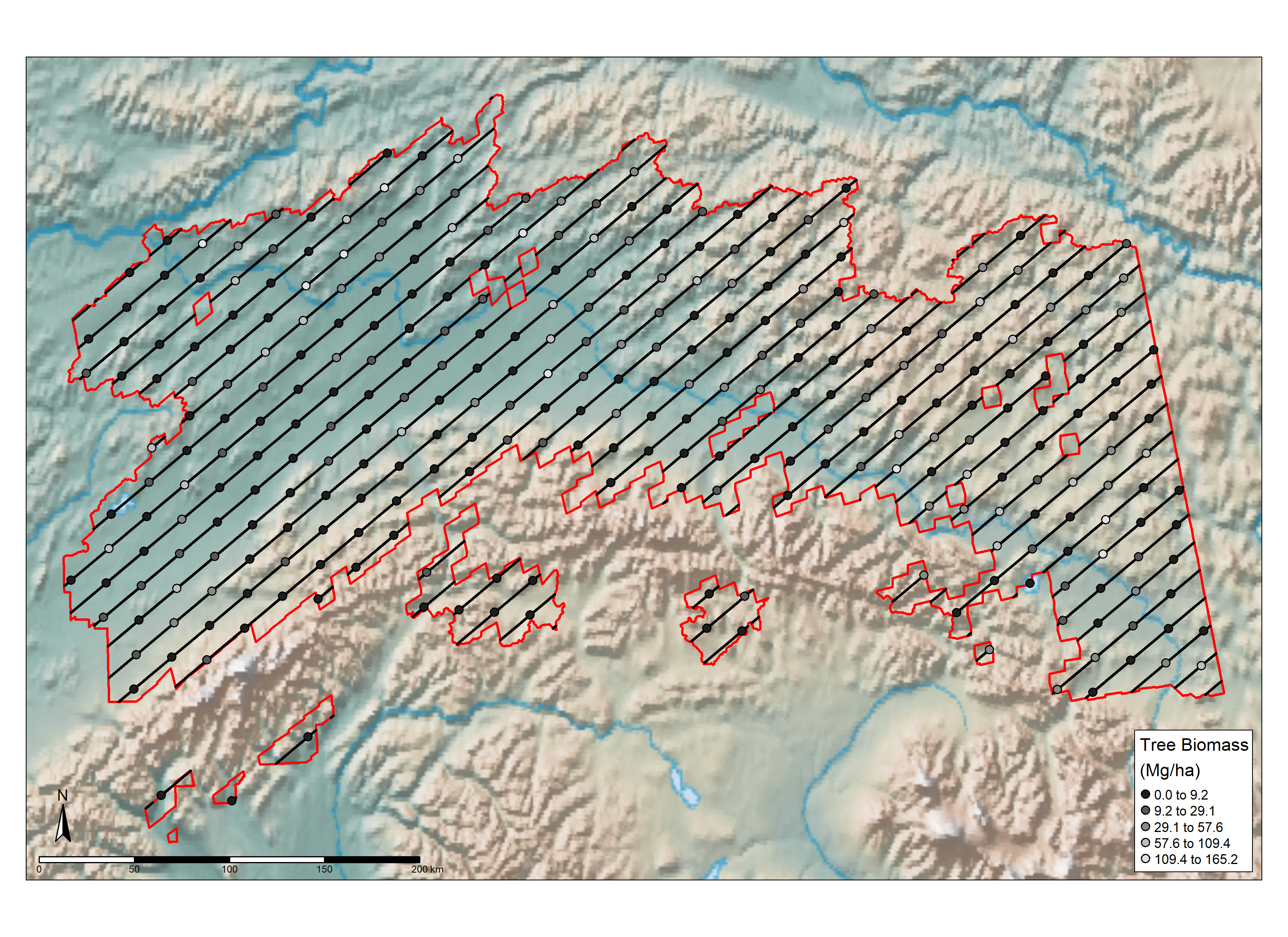}
    \caption{Field plot sampling intensity: 0.0075}
    \label{fig:second}
\end{subfigure}
\begin{subfigure}{0.6\textwidth}
    \includegraphics[width=\textwidth]{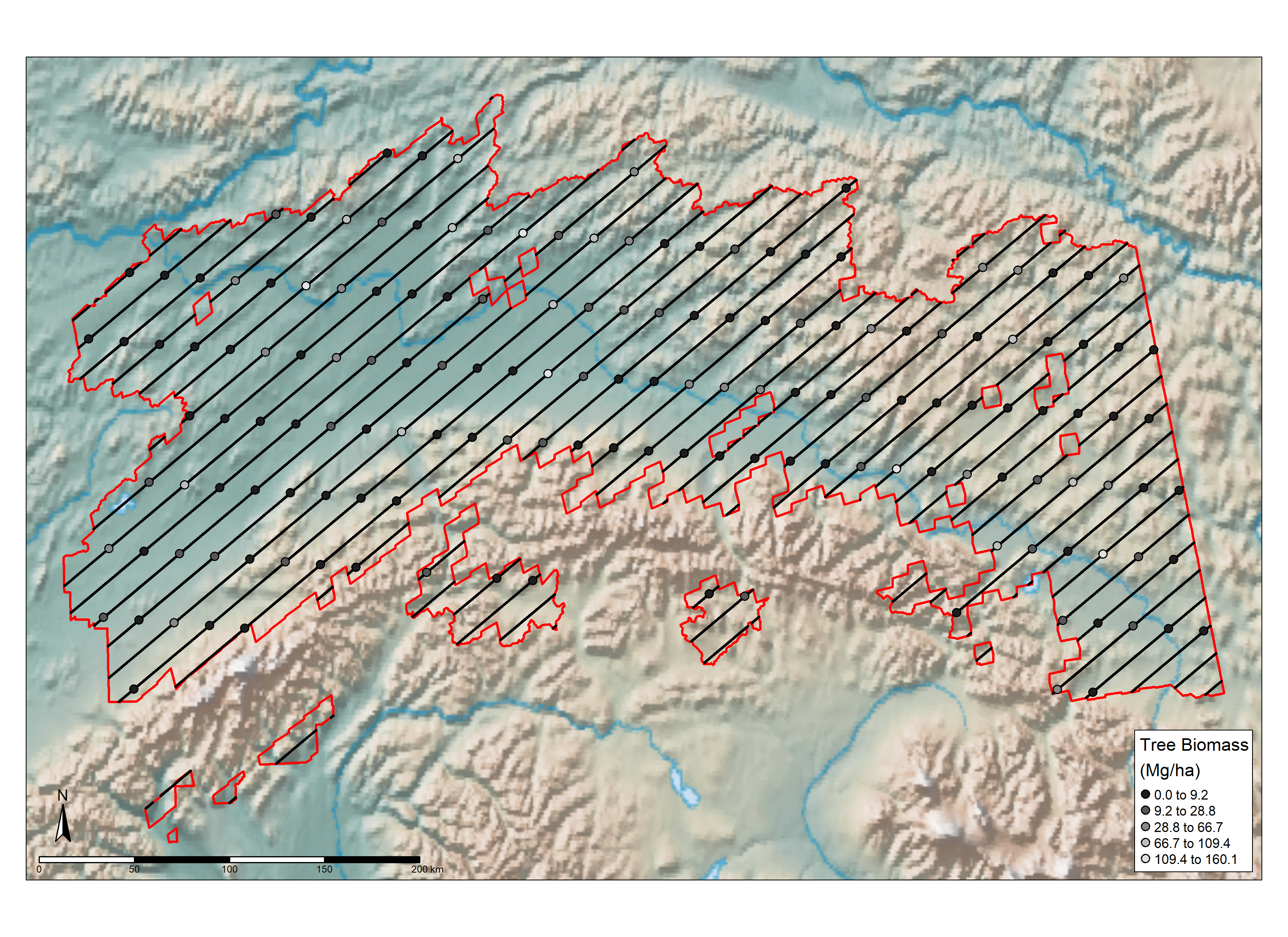}
    \caption{Field plot sampling intensity: 0.00375}
    \label{fig:third}
\end{subfigure}

\caption{Simulated samples at different sampling intensities with systematic sampling. Black lines indicate random sampled of airborne lidar flight lines. Dots indicate simulated field plots, color-coded by biomass}
\label{fig:figures}
\end{figure}

In this study, we use a combination of modeling approaches and
nearest-neighbor imputation to develop a simulated population over the
Tanana study area, with the objectives of 1) generating values for
independent and dependent variables that have realistic
\emph{marginal distributions}, \emph{correlation structure}, and
\emph{spatial distributions}. Using the methods described in (Ene et al.
(2012)), a Gaussian copula model was used to generate a large (200,000)
sample of simulated plots with marginal distributions and correlation
structure similar to the observed plot data in the Tanana study area. To
simulate the value for forestland proportion, we develop a logistic
regression model using the observed data with forestland class (forested
vs.~nonforested) as the dependent variable and lidar height as the
independent variable. For each of the 200,000 simulated values generated
from the above copula model, we draw a random value from a binomial
distribution where the parameter is the predicted probability from
logistic regression. To introduce realistic spatial variability in the
simulated population, we generate a dense grid of points (grid spacing
approximately 200 meters) across the entire Tanana study area (2,703,291
total points), and at each of these points attach a value for forest
cover (0-100\(%
\)) from a global forest cover classification layer (Hansen et al.
(2013)). Finally, a Most Similar Neighbor (MSN) imputation algorithm was
applied using the \texttt{yaimpute} package in \texttt{R} to assign plot
values (stratum, lidar height, tree biomass, forestland proportion) to
each point in the dense forest cover grid (Crookston and Finley (2007)).

In this way, the marginal distributions (Figure 2) and correlation
structure (Table 2) in the original plot data are generally maintained
in the simulated population, while association with mapped forest canopy
cover ensures that the simulated population is realistically distributed
in the spatial domain (Figure 3). In addition, each point in the dense
grid is assigned a strip number to enable selection of random strip
samples. This simulated population can then be used to select samples
and assess the statistical properties (bias, variability) of alternative
estimators.

At each iteration, a simple random sample of \(m\) \emph{PSU}s (lidar
strips) is drawn from the population, and from this a subsample of \(n\)
\emph{SSU}s (plots) are drawn as a simple random sample, at a specified
plot sampling intensity (defined as the ratio
\(\frac{\text{field plots}}{\text{remote sensing plots}}\)), similar to
the the approach used in Ringvall et al. (2016). This approach allows
for assessment of statistical properties for different sampling
intensities for field plots (in this study we assume that the sampling
intensity of the lidar is fixed) (Figure 4). In order to evaluate the
estimators under systematic sampling, simulated samples were drawn where
the airborne remote sensing strips were collected at regular intervals
and the field plots were distributed systematically within the airborne
strips (with the same number of strips and field sampling intensities as
the SRS sampling) (Figure 5).

\hypertarget{simulation-based-metrics}{%
\subsection{Simulation-based metrics}\label{simulation-based-metrics}}

The properties of each estimator are assessed by drawing a large number
(\emph{K}) of simulated samples from the population and using these
draws to develop a sampling distribution for each point estimator and
corresponding variance estimator, which can be used to evaluate the
bias, variance, as well as other measures such as the empirical coverage
probability. In this study, we used \(K\)=10,000, which was considered
to be an adequately large series in Särndal, Swensson, and Wretman
(2003). For example, for an estimator of an arbitrary population
parameter \(Y\) (i.e.~a total, area, or density), at each iteration, the
point estimator \(\widehat{Y_k}\) and the standard deviation estimator
\(\sqrt{\widehat{V}\left( \widehat{Y_k} \right)}\) are calculated based
on the equations presented in the previous sections, and the mean,
standard deviation, and observed bias (in percent) are calculated for
both the point estimator and standard deviation estimator based on the
\emph{K} simulated samples (see Ringvall et al. (2016) for details on
these calculations). In addition, the
\emph{empirical coverage probability of the} 95\%
\emph{confidence interval} for the point estimator was calculated as the
proportion of simulations where the calculated 95\% confidence
interval:\\
\(\left( \widehat{Y}-z_{0.025}\sqrt{\widehat{V}\left( \widehat{Y} \right)}, \widehat{Y}+z_{0.975}\sqrt{\widehat{V}\left( \widehat{Y} \right)} \right)\)
contains the true value \emph{Y}. The empirical coverage probability
provides an indication of how reliable the CI is as measure of
uncertainty. An empirical coverage probability near 0.95 is an indicator
that the 95\% confidence intervals calculated using this estimator are
reliable. Empirical coverage probabilities less than 0.95 indicate that
the calculated 95\% CIs are giving a falsely precise estimate of
uncertainty, while coverage probabilities greater than 0.95 indicate
that the 95\% CIs obtained from this estimator are overly conservative.

\hypertarget{results-and-discussion}{%
\subsection{Results and discussion}\label{results-and-discussion}}

\hypertarget{simulation-results}{%
\subsubsection{Simulation results}\label{simulation-results}}

\begin{table}[H]
    \begin{center}
    \caption{Simulation results, Simple Random Sampling, Biomass ($kt$)}
    \label{tab:table1}
    \begin{tabular}{ccccccccc}
        \toprule
        \multirow{2}{*}{$t$} & \multirow{2}{*}{Estimator} & \multicolumn{3}{c}{\textbf{$\widehat{t}$}} & \multicolumn{3}{c}{\textbf{$\sqrt{\widehat{V}(\widehat{t})}$}} & \multirow{2}{*}{\makecell{$95\%$ CI Cov.\\Prob.}} \\
        \cmidrule(lr){3-5} \cmidrule(lr){6-8} \\
        {} & {} & Mean & St.Dev. & Bias$(\%)$ & Mean & St.Dev. & Bias$(\%)$ & {} \\
        \midrule
       \multirow{15}{*}{222,621} & {} & \multicolumn{7}{c}{\textbf{Plot Sampling Intensity: 0.015}} \\
        {} & SRS,PS & 222,519 &
        10,333 &
        -0.046 & 10,201  & 598 & -1.3 & 0.945 \\
        {} & R & 222,629 &
        8,985 & 0.0036 & 8,878  & 1,033 & -1.2 & 0.942 \\
        {} & \makecell{R,PS} & 222,665 &
        6,982 & 0.02 & 6,392  & 812 & -8.4 & 0.928 \\
        \cmidrule(lr){3-9} \\
        {} & {} & \multicolumn{7}{c}{\textbf{Plot Sampling Intensity: 0.0075}} \\
        {} & SRS,PS & 222,296 &
        14,573 & -0.15 & 14,482  & 1,094 & -0.62 & 0.945 \\
        {} & R & 222,648 &
        11,341 & 0.012 & 11,112  & 1,353 & -2 & 0.937 \\
        {} & \makecell{R,PS} & 222,801 &
        10,035 & 0.081 & 8,277  & 1,218 & -18 & 0.908 \\
        \cmidrule(lr){3-9} \\
        {} & {} & \multicolumn{7}{c}{\textbf{Plot Sampling Intensity: 0.00375}} \\
        {} & SRS,PS & 222,476 &
        20,458 & -0.065 & 20,464  & 2,085 & 0.03 & 0.944 \\
        {} & R & 222,718 &
        14,837 & 0.044 & 14,432  & 1,964 & -2.7 & 0.933 \\
        {} & \makecell{R,PS} & 222,911 &
        13,806 & 0.13 & 9,681  & 1,661 & -30 & 0.828 \\
        \bottomrule
    \end{tabular}
  \end{center}
\end{table}

\begin{table}[H]
    \begin{center}
    \caption{Simulation results, Simple Random Sampling, Area of Forestland ($km^{2}$)}
    \label{tab:table1}
    \begin{tabular}{ccccccccc}
        \toprule
        \multirow{2}{*}{$A$} & \multirow{2}{*}{Estimator} & \multicolumn{3}{c}{\textbf{$\widehat{A}$}} & \multicolumn{3}{c}{\textbf{$\sqrt{\widehat{V}(\widehat{A})}$}} & \multirow{2}{*}{\makecell{$95\%$ CI Cov.\\Prob.}} \\
        \cmidrule(lr){3-5} \cmidrule(lr){6-8} \\
        {} & {} & Mean & St.Dev. & Bias$(\%)$ & Mean & St.Dev. & Bias$(\%)$ & {} \\
        \midrule
        \multirow{15}{*}{68,536} & {} & \multicolumn{7}{c}{\textbf{Plot Sampling Intensity: 0.015}} \\
        {} & SRS,PS & 68,518 &
        1,671 & -0.026 & 1,674  & 62 & 0.18 & 0.950 \\
        {} & R & 68,543 &
        1,508 & 0.0097 & 1,491  & 162 & -1.2 & 0.942 \\
        {} & \makecell{R,PS} & 68,540 &
        1,515 & 0.005 & 1,472  & 162 & -2.8 & 0.928 \\
        \cmidrule(lr){3-9} \\
        {} & {} & \multicolumn{7}{c}{\textbf{Plot Sampling Intensity: 0.0075}} \\
        {} & SRS,PS & 68,521 &
        2,378 & -0.022 & 2,374  & 92 & -0.16 & 0.950 \\
        {} & R & 68,518 &
        2,098 & -0.027 & 2,063  & 219 & -1.7 & 0.942 \\
        {} & \makecell{R,PS} & 68,507 &
        2,185 & -0.043 & 2,071  & 229 & -5.2 & 0.908 \\
        \cmidrule(lr){3-9} \\
        {} & {} & \multicolumn{7}{c}{\textbf{Plot Sampling Intensity: 0.00375}} \\
        {} & SRS,PS & 68,536 &
        3,329 & 0.00044 & 3,342  & 145 & 0.42 & 0.949 \\
        {} & R & 68,557 &
        2,915 & 0.03 & 2,868  & 306 & -1.6 & 0.940 \\
        {} & \makecell{R,PS} & 68,551 &
        3,098 & 0.021 & 2,741  & 315 & -12 & 0.914 \\
        \bottomrule
    \end{tabular}
  \end{center}
\end{table}

\begin{table}[H]
    \begin{center}
    \caption{Simulation results, Simple Random Sampling, Average Biomass in Forestland ($Mg/ha$)}
    \label{tab:table1}
    \begin{tabular}{ccccccccc}
        \toprule
        \multirow{2}{*}{$D$} & \multirow{2}{*}{Estimator} & \multicolumn{3}{c}{\textbf{$\widehat{D}$}} & \multicolumn{3}{c}{\textbf{$\sqrt{\widehat{V}(\widehat{D})}$}} & \multirow{2}{*}{\makecell{$95\%$ CI Cov.\\Prob.}} \\
        \cmidrule(lr){3-5} \cmidrule(lr){6-8} \\
        {} & {} & Mean & St.Dev. & Bias$(\%)$ & Mean & St.Dev. & Bias$(\%)$ & {} \\
        \midrule
        \multirow{15}{*}{32.48} & {} & \multicolumn{7}{c}{\textbf{Plot Sampling Intensity: 0.015}} \\
        {} & SRS,PS & 32.49 &
        1.50 & 0.0094 & 1.49  & 0.091 & -0.65 & 0.945 \\
        {} & RoR & 32.49 &
        1.38 & 0.027 & 1.36  & 0.16 & -1.4 & 0.943 \\
        {} & \makecell{RoR,PS} & 32.50 &
        1.24 & 0.063 & 1.16  & 0.14 & -6.6 & 0.928 \\
        \cmidrule(lr){3-9} \\
        {} & {} & \multicolumn{7}{c}{\textbf{Plot Sampling Intensity: 0.0075}} \\
        {} & SRS,PS & 32.46 &
        2.16 & -0.058 & 2.12  & 0.17 & -1.7 & 0.944 \\
        {} & RoR & 32.52 &
        1.84 & 0.12 & 1.81  & 0.22 & -1.8 & 0.941 \\
        {} & \makecell{RoR,PS} & 32.56 &
        1.79 & 0.22 & 1.56  & 0.21 & -13 & 0.908 \\
        \cmidrule(lr){3-9} \\
        {} & {} & \multicolumn{7}{c}{\textbf{Plot Sampling Intensity: 0.00375}} \\
        {} & SRS,PS & 32.50 &
        2.99 & 0.054 & 3.00  & 0.32 & 0.12 & 0.948 \\
        {} & RoR & 32.54 &
        2.50 & 0.18 & 2.46  & 0.33 & -1.5 & 0.940 \\
        {} & \makecell{RoR,PS} & 32.58 &
        2.50 & 0.31 & 1.93  & 0.29 & -23 & 0.861 \\
        \bottomrule
    \end{tabular}
  \end{center}
\end{table}

\begin{table}[H]
    \begin{center}
    \caption{Simulation results, Systematic Sampling, Biomass ($kt$)}
    \label{tab:table1}
    \begin{tabular}{ccccccccc}
        \toprule
        \multirow{2}{*}{$t$} & \multirow{2}{*}{Estimator} & \multicolumn{3}{c}{\textbf{$\widehat{t}$}} & \multicolumn{3}{c}{\textbf{$\sqrt{\widehat{V}(\widehat{t})}$}} & \multirow{2}{*}{\makecell{$95\%$ CI Cov.\\Prob.}} \\
        \cmidrule(lr){3-5} \cmidrule(lr){6-8} \\
        {} & {} & Mean & St.Dev. & Bias$(\%)$ & Mean & St.Dev. & Bias$(\%)$ & {} \\
        \midrule
       \multirow{15}{*}{222,621} & {} & \multicolumn{7}{c}{\textbf{Plot Sampling Intensity: 0.015}} \\
        {} & SRS,PS & 222,633 &
        11,773 &
        0.0056 & 11,973  & 656 & 1.7 & 0.954 \\
        {} & R & 222,694 &
        8,121 & 0.033 & 10,476  & 1,255 & 29 & 0.985 \\
        {} & \makecell{R,PS} & 222,682 &
        8,027 & 0.027 & 7,424  & 1,058 & -7.5 & 0.924 \\
        \cmidrule(lr){3-9} \\
        {} & {} & \multicolumn{7}{c}{\textbf{Plot Sampling Intensity: 0.0077}} \\
        {} & SRS,PS & 222,641 &
        16,878 & 0.0092 & 17,000  & 1,323 & 0.72 & 0.949 \\
        {} & R & 222,767 &
        11,339 & 0.066 & 13,026  & 1,742 & 15 & 0.967 \\
        {} & \makecell{R,PS} & 222,817 &
        11,474 & 0.088 & 9,704  & 1,568 & -15 & 0.912 \\
        \cmidrule(lr){3-9} \\
        {} & {} & \multicolumn{7}{c}{\textbf{Plot Sampling Intensity: 0.00385}} \\
        {} & SRS,PS & 222,465 &
        24,481 & -0.07 & 24,190  & 2,781 & -1.2 & 0.942 \\
        {} & R & 222,886 &
        16,250 & 0.12 & 17,019  & 2,632 & 4.7 & 0.949 \\
        {} & \makecell{R,PS} & 223,006 &
        16,205 & 0.17 & 11,521  & 2,255 & -29 & 0.824 \\
        \bottomrule
    \end{tabular}
  \end{center}
\end{table}

\begin{table}[H]
    \begin{center}
    \caption{Simulation results, Systematic Sampling, Area of Forestland ($km^{2}$)}
    \label{tab:table1}
    \begin{tabular}{ccccccccc}
        \toprule
        \multirow{2}{*}{$A$} & \multirow{2}{*}{Estimator} & \multicolumn{3}{c}{\textbf{$\widehat{A}$}} & \multicolumn{3}{c}{\textbf{$\sqrt{\widehat{V}(\widehat{A})}$}} & \multirow{2}{*}{\makecell{$95\%$ CI Cov.\\Prob.}} \\
        \cmidrule(lr){3-5} \cmidrule(lr){6-8} \\
        {} & {} & Mean & St.Dev. & Bias$(\%)$ & Mean & St.Dev. & Bias$(\%)$ & {} \\
        \midrule
        \multirow{15}{*}{68,536} & {} & \multicolumn{7}{c}{\textbf{Plot Sampling Intensity: 0.015}} \\
        {} & SRS,PS & 68,537 &
        1,978 & 0.001 & 1,963  & 29 & -0.73 & 0.947 \\
        {} & R & 68,536 &
        1,702 & 0.0000028 & 1,748  & 211 & 2.7 & 0.953 \\
        {} & \makecell{R,PS} & 68,542 &
        1,763 & 0.0088 & 1,705  & 207 & -3.3 & 0.924 \\
        \cmidrule(lr){3-9} \\
        {} & {} & \multicolumn{7}{c}{\textbf{Plot Sampling Intensity: 0.0077}} \\
        {} & SRS,PS & 68,536 &
        2,806 & 0.0002 & 2,781  & 61 & -0.92 & 0.947 \\
        {} & R & 68,534 &
        2,417 & -0.0031 & 2,411  & 290 & -0.22 & 0.942 \\
        {} & \makecell{R,PS} & 68,536 &
        2,544 & -0.00081 & 2,403  & 298 & -5.5 & 0.912 \\
        \cmidrule(lr){3-9} \\
        {} & {} & \multicolumn{7}{c}{\textbf{Plot Sampling Intensity: 0.00385}} \\
        {} & SRS,PS & 68,513 &
        3,988 & -0.034 & 3,943  & 130 & -1.1 & 0.942 \\
        {} & R & 68,532 &
        3,452 & -0.0063 & 3,390  & 406 & -1.8 & 0.936 \\
        {} & \makecell{R,PS} & 68,521 &
        3,656 & -0.022 & 3,222  & 421 & -12 & 0.908 \\
        \bottomrule
    \end{tabular}
  \end{center}
\end{table}

\begin{table}[H]
    \begin{center}
    \caption{Simulation results, Systematic Sampling, Average Biomass in Forestland ($Mg/ha$)}
    \label{tab:table1}
    \begin{tabular}{ccccccccc}
        \toprule
        \multirow{2}{*}{$D$} & \multirow{2}{*}{Estimator} & \multicolumn{3}{c}{\textbf{$\widehat{D}$}} & \multicolumn{3}{c}{\textbf{$\sqrt{\widehat{V}(\widehat{D})}$}} & \multirow{2}{*}{\makecell{$95\%$ CI Cov.\\Prob.}} \\
        \cmidrule(lr){3-5} \cmidrule(lr){6-8} \\
        {} & {} & Mean & St.Dev. & Bias$(\%)$ & Mean & St.Dev. & Bias$(\%)$ & {} \\
        \midrule
        \multirow{15}{*}{32.48} & {} & \multicolumn{7}{c}{\textbf{Plot Sampling Intensity: 0.015}} \\
        {} & SRS,PS & 32.50 &
        1.72 & 0.047 & 1.75  & 0.10 & 1.6 & 0.953 \\
        {} & RoR & 32.51 &
        1.41 & 0.091 & 1.60  & 0.21 & 13 & 0.966 \\
        {} & \makecell{RoR,PS} & 32.51 &
        1.42 & 0.082 & 1.35  & 0.19 & -5.1 & 0.924 \\
        \cmidrule(lr){3-9} \\
        {} & {} & \multicolumn{7}{c}{\textbf{Plot Sampling Intensity: 0.0077}} \\
        {} & SRS,PS & 32.51 &
        2.48 & 0.096 & 2.49  & 0.21 & 0.29 & 0.947 \\
        {} & RoR & 32.54 &
        2.01 & 0.19 & 2.11  & 0.29 & 5.4 & 0.958 \\
        {} & \makecell{RoR,PS} & 32.56 &
        2.06 & 0.22 & 1.82  & 0.27 & -12 & 0.912 \\
        \cmidrule(lr){3-9} \\
        {} & {} & \multicolumn{7}{c}{\textbf{Plot Sampling Intensity: 0.00385}} \\
        {} & SRS,PS & 32.53 &
        3.59 & 0.13 & 3.55  & 0.44 & -1.1 & 0.943 \\
        {} & RoR & 32.61 &
        2.89 & 0.38 & 2.91  & 0.45 & 0.67 & 0.942 \\
        {} & \makecell{RoR,PS} & 32.64 &
        2.95 & 0.48 & 2.29  & 0.40 & -22 & 0.863 \\
        \bottomrule
    \end{tabular}
  \end{center}
\end{table}

The ratio estimators for total biomass (\(\widehat{t}_{R}\) and
\(\widehat{t}_{R, PS}\)), forestland area ((\(\widehat{A}_{R}\) and
\(\widehat{A}_{R, PS}\)), and biomass density in forestland
(\(\widehat{D}_{RoR}\) and \(\widehat{D}_{RoR, PS}\)) were generally
unbiased at all sampling intensities and sampling configurations (simple
random sampling vs.~systematic), as was the standard poststratified
estimator (\(\widehat{t}_{SRS,PS}\), \(\widehat{A}_{SRS,PS}\), and
\(\widehat{D}_{SRS,PS}\)) (Tables 3-5). While the observed bias was low
(\textless{} 0.34\%) for all estimators, it was higher for ratio
estimators than the standard poststratified estimator. It should be
noted that model-assisted estimators, such as the regression estimator
that is a component of the ratio estimators used here, are
\emph{approximately} design-unbiased, which means that they may be
biased for small sample sizes (Särndal, Swensson, and Wretman (2003)).
Furthermore, in the case of poststratified, two-stage sampling designs,
one must consider the sample sizes by \emph{stratum} and \emph{PSU},
which can be quite small, potentially introducing bias in the
estimators.

The ratio estimators of total biomass (Table 3) were significantly more
precise than the standard poststratified estimator
\(\sqrt{V(\widehat{t}_{SRS,PS})}\), with a reduction of approx. 31\% for
\(\sqrt{V(\widehat{t}_{R,PS})}\) and 11\% for
\(\sqrt{V(\widehat{t}_{R})}\) at the 0.015 sampling intensity, which
also indicates the benefit of using poststratification. Ratio estimators
of forestland area (Table 4) were more precise than the standard
poststratified estimator, but the gains were modest and, interestingly,
poststratification did not improve the precision of the ratio estimator
(11\% reduction for \(\sqrt{V(\widehat{A}_{R})}\) in relation to
\(\sqrt{V(\widehat{A}_{SRS,PS})}\), but only 10\% reduction for
\(\sqrt{V(\widehat{A}_{R, PS})}\) at 0.015 sampling intensity). There
were moderate gains in precision from using the ratio estimators in
estimation of average tree biomass in forestland (Table 5) (16\%
reduction for \(\sqrt{V(\widehat{D}_{RoR,PS})}\) in relation to
\(\sqrt{V(\widehat{D}_{SRS,PS})}\), 8\% reduction for
\(\sqrt{V(\widehat{D}_{RoR})}\) at the 0.015 sampling intensity).

The relative precision of the ratio estimators (vs.~standard
poststratified estimators) is more pronounced as the field sampling
intensity decreased. For example, at the lowest sampling intensity of
0.0375, the reduction in standard error (relative to standard
poststratified estimators) for ratio estimators was 27\% for
\(\widehat{t}_{R}\) and 33\% for \(\widehat{t}_{R,PS}\), 11\% for
\(\widehat{A}_{R}\) and 6\% for \(\widehat{A}_{R,PS}\), and 16\% for
\(\widehat{D}_{RoR}\) and 17\% for \(\widehat{D}_{RoR,PS}\).

The SE estimators in the standard FIA poststratification estimation
framework are generally unbiased (i.e.~bias \textless{} 1\%) for all
attributes and at all sampling intensities. The observed bias of SE
estimators for the ratio estimators without poststratification was low,
while PS ratio estimators are increasingly biased at low sampling
intensities. The problem of negative bias in the variance of ratio
estimators with smaller sample sizes has been reported in previous
studies (Knottnerus and Scholtus (2019)), and the further development of
these estimators to reduce this bias should be a focus of future
research.

The use of variance estimators that assume simple random sampling, when
the sampling is actually systematic results in higher estimates of
standard error (Tables 6-8). This result is often observed when sampling
natural populations (such as forests) where attributes are spatially
correlated, because systematic samples force a minimum distance between
samples, whereas simple random sampling does not (Gregoire and Valentine
(2007); Cochran (1977)). Although the observed effect of this
overestimation of SE in this study is generally small, it does result in
several cases (notably at higher field sampling intensities) where 95\%
CI coverage probabilities for \(SRS,PS\), \(R\), and \(RoR\) estimators
exceed 0.95. The use of SRS variance estimators with systematic samples
is usually considered to be acceptable in NFI programs because it
typically results in a conservative estimate for precision (i.e.,
overestimate of SE).

In the context of forest inventory in remote regions, it can be useful
to express gains in precision by using an alternative design in terms of
the increase in plot sample size needed to achieve this precision under
the standard design. For example, to achieve the 31\% reduction in
standard error from using the \(\widehat{t}_{R,PS}\) estimator would
require 2.1 times more field plots in the standard FIA sampling design
and estimation framework. Even modest reductions in standard error can
lead to significant cost savings; for example, a reduction in standard
error of 10\% for \(\widehat{A}_{R,PS}\) equates to using 1.23 times
more plots in the standard design, a significant cost savings when cost
per plot is high (e.g., \textasciitilde{} \$10,000 USD/plot in interior
Alaska) and annual number of plots installed exceeds one hundred (Cahoon
and Baer (2022)). Although the G-LiHT airborne data was collected as
part of a research study and it is therefore difficult to estimate the
total cost of data collection, if the total cost of the G-LiHT
acquisition (including transit, salaries, etc.) is conservatively
estimated at \$50/km, this represents a fraction of the cost of the the
increased field plot sample required to match the statistical precision
of the two-stage design.

It should be noted that although the form of the variance estimators can
appear complex, the fundamental principles behind both two-stage
sampling and regression estimation still apply. For example, the
precision of the model-assisted ratio estimators in a two-stage design
will depend on how evenly the domain of interest occurs across
\emph{PSU}s. With a large variation between \emph{PSU}s, the
\(s_{r}^{2}\) (or \(s_{rh}^{2}\)) term would also most likely be large.
This, however, does not necessarily mean that the ratio-of-ratios
estimator of the domain density has a large variance. Also, it is
apparent that improvement in the models to predict \(y_{i}\) and
\(a_{i}\), will reduce the error terms \(s_{e}^{2}\), leading to a
smaller variance for the estimators.

In conclusion, the methods and results presented here indicate that
airborne remote sensing data -- collected in a sampling mode -- can
increase the efficiency of national forest inventories, especially in
remote regions. The ratio-of-ratio estimator developed in this study
enables estimation of population densities at the domain level within
the context of the two-stage sampling design, and together with the
ratio estimators previously presented in Ringvall et al. (2016), provide
a full suite of model-assisted estimators that can serve as alternatives
to the standard poststratified estimators in Bechtold and Patterson
(2005) when an intermediate level of sampled airborne data is available.
Model-assisted estimation, which includes both standard
post-stratification and regression-based approaches, supports
design-unbiased estimation of inventory attributes, and associated
confidence intervals, and therefore is well-suited to the NFI context,
where a nationally-consistent estimation framework is required.

It should be noted that the focus of this study was on the estimation
framework, and not on the challenges of classifying domains or
prediction of inventory attributes via airborne remote sensing. It is
expected that future efforts will be devoted to the development of more
sophisticated models for domain (e.g.,forest type) and attribute
characterization, which could significantly increase the precision of
estimators presented here. For example, the use of machine learning
algorithms (presumably trained on a large, external (i.e.~independent)
datasets to avoid overfitting), and applied to high-resolution,
hyperspectral/hyperspatial imagery has the potential to markedly improve
domain classification and forest attribute prediction (Dalponte et al.
(2014); Breidt and Opsomer (2017)). In addition, the estimators
presented here focus on inventory attributes associated with forest
\emph{status}, and not \emph{trends}. While several recent studies have
reported that a similar two-stage sampling design and model-assisted
estimation framework can be extended to estimate of change (Strîmbu et
al. (2017)), further research is required to develop and evaluate the
full set of change estimators required in a NFI context.

\hypertarget{acknowledgments}{%
\section{Acknowledgments}\label{acknowledgments}}

Funding for this project was provided by the USDA Forest Service Pacific
Northwest Research Station and the National Aeronautics and Space
Administration (NASA). This research was part of the project ``A Joint
USFS-NASA Pilot Project to Estimate Forest Carbon Stocks in Interior
Alaska by Integrating Field, Airborne and Satellite Data'' funded by the
NASA Carbon Monitoring System (CMS) (NNX13AQ51G). We would like to thank
the interior Alaska FIA field crews of the Alaska Department of Natural
Resources, Division of Forestry for high-quality data collection under
extremely challenging conditions, and GIS support from Thomas Thompson
of the USDA Forest Service, Anchorage Forestry Sciences Laboratory. The
authors would also like to thank Larry Corp and Ross Nelson for their
significant contributions in the design, planning, and implementation of
this study, and Tad Fickel for his participation in the 2014 G-LiHT
campaign.

\hypertarget{references}{%
\section{References}\label{references}}

\hypertarget{refs}{}
\begin{CSLReferences}{1}{0}
\leavevmode\vadjust pre{\hypertarget{ref-andersen2009using}{}}%
Andersen, Hans-Erik. 2009. {``Using Airborne Light Detection and Ranging
(LIDAR) to Characterize Forest Stand Condition on the Kenai Peninsula of
Alaska.''} \emph{Western Journal of Applied Forestry} 24 (2): 95--102.

\leavevmode\vadjust pre{\hypertarget{ref-andersen2009estimating}{}}%
Andersen, Hans-Erik, Tara Barrett, Ken Winterberger, Jacob Strunk, and
Hailemariam Temesgen. 2009. {``Estimating Forest Biomass on the Western
Lowlands of the Kenai Peninsula of Alaska Using Airborne Lidar and Field
Plot Data in a Model-Assisted Sampling Design.''} In \emph{Proceedings
of the IUFRO Division 4 Conference:``extending Forest Inventory and
Monitoring over Space and Time}, 19--22.

\leavevmode\vadjust pre{\hypertarget{ref-andersen2009accuracy}{}}%
Andersen, Hans-Erik, Tobey Clarkin, Ken Winterberger, and Jacob Strunk.
2009. {``An Accuracy Assessment of Positions Obtained Using Survey-and
Recreational-Grade Global Positioning System Receivers Across a Range of
Forest Conditions Within the Tanana Valley of Interior Alaska.''}
\emph{Western Journal of Applied Forestry} 24 (3): 128--36.

\leavevmode\vadjust pre{\hypertarget{ref-andersen2022using}{}}%
Andersen, Hans-Erik, Jacob Strunk, and Robert J McGaughey. 2022.
\emph{Using High-Performance Global Navigation Satellite System
Technology to Improve {Forest Inventory and Analysis} Plot Coordinates
in the Pacific Region}. PNW-GTR-1000. USDA Forest Service, Pacific
Northwest Research Station.

\leavevmode\vadjust pre{\hypertarget{ref-babcock2018geostatistical}{}}%
Babcock, Chad, Andrew O Finley, Hans-Erik Andersen, Robert Pattison,
Bruce D Cook, Douglas C Morton, Michael Alonzo, et al. 2018.
{``Geostatistical Estimation of Forest Biomass in Interior Alaska
Combining Landsat-Derived Tree Cover, Sampled Airborne Lidar and Field
Observations.''} \emph{Remote Sensing of Environment} 212: 212--30.

\leavevmode\vadjust pre{\hypertarget{ref-bechtold2005enhanced}{}}%
Bechtold, William A, and Paul L Patterson. 2005. \emph{The Enhanced
Forest Inventory and Analysis Program--National Sampling Design and
Estimation Procedures}. SRS-GTR-80. USDA Forest Service, Southern
Research Station.

\leavevmode\vadjust pre{\hypertarget{ref-breidt2017model}{}}%
Breidt, F Jay, and Jean D Opsomer. 2017. {``Model-Assisted Survey
Estimation with Modern Prediction Techniques.''} \emph{Statistical
Science} 32 (2): 190--205.

\leavevmode\vadjust pre{\hypertarget{ref-cahoon2022tanana}{}}%
Cahoon, Sean, and Kathryn Baer. 2022. \emph{Forest Resources of the
Tanana Unit, Alaska: 2018}. PNW-GTR-1005. USDA Forest Service, Pacific
Northwest Research Station.

\leavevmode\vadjust pre{\hypertarget{ref-cochran1977sampling}{}}%
Cochran, William. 1977. \emph{Sampling Techniques}. Wiley, New York.

\leavevmode\vadjust pre{\hypertarget{ref-cook2013nasa}{}}%
Cook, Bruce D, Lawrence A Corp, Ross F Nelson, Elizabeth M Middleton,
Douglas C Morton, Joel T McCorkel, Jeffrey G Masek, Kenneth J Ranson,
Vuong Ly, and Paul M Montesano. 2013. {``NASA Goddard's LiDAR,
Hyperspectral and Thermal ({G-LiHT}) Airborne Imager.''} \emph{Remote
Sensing} 5 (8): 4045--66.

\leavevmode\vadjust pre{\hypertarget{ref-crookstonfinley2007}{}}%
Crookston, Nicholas L., and Andrew O. Finley. 2007. {``yaImpute: An {R}
Package for kNN Imputation.''} \emph{Journal of Statistical Software} 23
(10). \url{https://doi.org/10.18637/jss.v023.i10}.

\leavevmode\vadjust pre{\hypertarget{ref-dalponte2014treespecieshyperspec}{}}%
Dalponte, Michele, Hans Ole Ørka, Liviu Theodor Ene, Terje Gobakken, and
Erik Næsset. 2014. {``Tree Crown Delineation and Tree Species
Classification in Boreal Forests Using Hyperspectral and ALS Data.''}
\emph{Remote Sensing of Environment} 140: 306--17.
https://doi.org/\url{https://doi.org/10.1016/j.rse.2013.09.006}.

\leavevmode\vadjust pre{\hypertarget{ref-Dettmann2022}{}}%
Dettmann, Garret T., Philip J. Radtke, John W. Coulston, P. Corey Green,
Barry T. Wilson, and Gretchen G. Moisen. 2022. {``Review and Synthesis
of Estimation Strategies to Meet Small Area Needs in Forest
Inventory.''} \emph{Frontiers in Forests and Global Change} 5 (March).
\url{https://doi.org/10.3389/ffgc.2022.813569}.

\leavevmode\vadjust pre{\hypertarget{ref-ene2018large}{}}%
Ene, Liviu, Terje Gobakken, Hans-Erik Andersen, Erik Næsset, Bruce D
Cook, Douglas C Morton, Chad Babcock, and Ross Nelson. 2018.
{``Large-Area Hybrid Estimation of Aboveground Biomass in Interior
Alaska Using Airborne Laser Scanning Data.''} \emph{Remote Sensing of
Environment} 204: 741--55.

\leavevmode\vadjust pre{\hypertarget{ref-ene2012assessing}{}}%
Ene, Liviu, Erik Næsset, Terje Gobakken, Timothy Gregoire, Göran Ståhl,
and Ross Nelson. 2012. {``Assessing the Accuracy of Regional LiDAR-Based
Biomass Estimation Using a Simulation Approach.''} \emph{Remote Sensing
of Environment} 123 (August): 579--92.
\url{https://doi.org/10.1016/j.rse.2012.04.017}.

\leavevmode\vadjust pre{\hypertarget{ref-epa2023}{}}%
EPA. 2023. {``Inventory of {U.S.} Greenhouse Gas Emissions and Sinks:
1990-2021.''} U.S. Environmental Protection Agency. 2023.
\url{https://www.epa.gov/ghgemissions/inventory-us-greenhouse-gas-emissions-and-sinks-1990-2021}.

\leavevmode\vadjust pre{\hypertarget{ref-fao2016}{}}%
FAO. 2016. {``Global Forest Resources Assessment 2015: How Are the
World's Forests Changing?''} United Nations Food; Agricultural
Organization. \url{https://www.fao.org/3/i4793e/i4793e.pdf}.

\leavevmode\vadjust pre{\hypertarget{ref-gobakken2012estimatingbiomass}{}}%
Gobakken, Terje, Erik Næsset, Ross Nelson, Ole Martin Bollandsås,
Timothy G. Gregoire, Göran Ståhl, Sören Holm, Hans Ole Ørka, and Rasmus
Astrup. 2012. {``Estimating Biomass in Hedmark County, Norway Using
National Forest Inventory Field Plots and Airborne Laser Scanning.''}
\emph{Remote Sensing of Environment} 123: 443--56.
https://doi.org/\url{https://doi.org/10.1016/j.rse.2012.01.025}.

\leavevmode\vadjust pre{\hypertarget{ref-gregoire2011model}{}}%
Gregoire, Timothy, Göran Ståhl, Erik Næsset, Terje Gobakken, Ross
Nelson, and Sören Holm. 2011. {``Model-Assisted Estimation of Biomass in
a LiDAR Sample Survey in Hedmark County, Norway.''} \emph{Canadian
Journal of Forest Research} 41 (1): 83--95.

\leavevmode\vadjust pre{\hypertarget{ref-gregoire2007sampling}{}}%
Gregoire, Timothy, and Harry Valentine. 2007. \emph{Sampling Strategies
for Natural Resources and the Environment}. Chapman; Hall/CRS, New York.

\leavevmode\vadjust pre{\hypertarget{ref-hansen2013globalforestcover}{}}%
Hansen, M. C., Peter Potapov, Rebecca Moore, M Hancher, Svetlana
Turubanova, Alexandra Tyukavina, D Thau, et al. 2013. {``High-Resolution
Global Maps of 21st-Century Forest Cover Change.''} \emph{Science (New
York, N.Y.)} 342 (November): 850--53.
\url{https://doi.org/10.1126/science.1244693}.

\leavevmode\vadjust pre{\hypertarget{ref-homer2015completion}{}}%
Homer, Collin, Jon Dewitz, Limin Yang, Suming Jin, Patrick Danielson,
George Xian, John Coulston, Nathaniel Herold, James Wickham, and Kevin
Megown. 2015. {``Completion of the 2011 National Land Cover Database for
the Conterminous United States--Representing a Decade of Land Cover
Change Information.''} \emph{Photogrammetric Engineering \& Remote
Sensing} 81 (5): 345--54.

\leavevmode\vadjust pre{\hypertarget{ref-kangasremotesensingnfis2018}{}}%
Kangas, Annika, Rasmus Astrup, Johannes Breidenbach, Jonas Fridman,
Terje Gobakken, Kari Korhonen, Matti Maltamo, et al. 2018. {``Remote
Sensing and Forest Inventories in Nordic Countries -- Roadmap for the
Future.''} \emph{Scandinavian Journal of Forest Research} 33 (January).
\url{https://doi.org/10.1080/02827581.2017.1416666}.

\leavevmode\vadjust pre{\hypertarget{ref-knotterusscholtus2019biasratioestimator}{}}%
Knottnerus, P., and S. Scholtus. 2019. {``On a New Estimator for the
Variance of the Ratio Estimator with Small Sample Corrections.''}
\emph{Survey Methodology} 45 (3): 567--76.
\url{https://www150.statcan.gc.ca/n1/pub/12-001-x/2019003/article/00003-eng.htm}.

\leavevmode\vadjust pre{\hypertarget{ref-lister2020use}{}}%
Lister, Andrew J, Hans Andersen, Tracey Frescino, Demetrios Gatziolis,
Sean Healey, Linda S Heath, Greg C Liknes, et al. 2020. {``Use of Remote
Sensing Data to Improve the Efficiency of National Forest Inventories: A
Case Study from the United States National Forest Inventory.''}
\emph{Forests} 11 (12): 1364.

\leavevmode\vadjust pre{\hypertarget{ref-magnussen2018lidar}{}}%
Magnussen, Steen, Thomas Nord-Larsen, and Torben Riis-Nielsen. 2018.
{``Lidar Supported Estimators of Wood Volume and Aboveground Biomass
from the Danish National Forest Inventory (2012--2016).''} \emph{Remote
Sensing of Environment} 211: 146--53.

\leavevmode\vadjust pre{\hypertarget{ref-marland2017}{}}%
Marland, Eric, Grant Domke, Jason Hoyle, Gregg Marland, Laurel Bates,
Alex Helms, Benjamin Jones, Tamara Kowalczyk, Tatyana B. Ruseva, and
Celina Szymanski. 2017. \emph{Understanding and Analysis: The California
Air Resources Board Forest Offset Protocol}. Springer Cham.
https://doi.org/\url{https://doi.org/10.1007/978-3-319-52434-4}.

\leavevmode\vadjust pre{\hypertarget{ref-mcconville2017model}{}}%
McConville, Kelly S, F Jay Breidt, Thomas Lee, and Gretchen G Moisen.
2017. {``Model-Assisted Survey Regression Estimation with the Lasso.''}
\emph{Journal of Survey Statistics and Methodology} 5 (2): 131--58.

\leavevmode\vadjust pre{\hypertarget{ref-mcgaughey2017effect}{}}%
McGaughey, Robert J, Kamal Ahmed, Hans-Erik Andersen, and Stephen E
Reutebuch. 2017. {``Effect of Occupation Time on the Horizontal Accuracy
of a Mapping-Grade GNSS Receiver Under Dense Forest Canopy.''}
\emph{Photogrammetric Engineering \& Remote Sensing} 83 (12): 861--68.

\leavevmode\vadjust pre{\hypertarget{ref-mcroberts2010probability}{}}%
McRoberts, Ronald E. 2010. {``Probability-and Model-Based Approaches to
Inference for Proportion Forest Using Satellite Imagery as Ancillary
Data.''} \emph{Remote Sensing of Environment} 114 (5): 1017--25.

\leavevmode\vadjust pre{\hypertarget{ref-mcroberts2014using}{}}%
McRoberts, Ronald E, Hans-Erik Andersen, and Erik Næsset. 2014. {``Using
Airborne Laser Scanning Data to Support Forest Sample Surveys.''} In
\emph{Forestry Applications of Airborne Laser Scanning}, 269--92.
Springer, Dordrecht.

\leavevmode\vadjust pre{\hypertarget{ref-ringvall2016poststratified}{}}%
Ringvall, Anna, Goran Stahl, Liviu Ene, Erik Næsset, Terje Gobakken, and
Timothy Gregoire. 2016. {``A Poststratified Ratio Estimator for
Model-Assisted Biomass Estimation in Sample-Based Airborne Laser
Scanning Surveys.''} \emph{Canadian Journal of Forest Research} 46
(July): 1386--95. \url{https://doi.org/10.1139/cjfr-2016-0158}.

\leavevmode\vadjust pre{\hypertarget{ref-saarela2017new}{}}%
Saarela, Svetlana, Hans-Erik Andersen, Anton Grafström, Sebastian
Schnell, Terje Gobakken, Erik Næsset, Ross F Nelson, Ronald E McRoberts,
Timothy G Gregoire, and Göran Ståhl. 2017. {``A New Prediction-Based
Variance Estimator for Two-Stage Model-Assisted Surveys of Forest
Resources.''} \emph{Remote Sensing of Environment} 192: 1--11.

\leavevmode\vadjust pre{\hypertarget{ref-saarela2016hierarchical}{}}%
Saarela, Svetlana, Sören Holm, Anton Grafström, Sebastian Schnell, Erik
Næsset, Timothy G Gregoire, Ross F Nelson, and Göran Ståhl. 2016.
{``Hierarchical Model-Based Inference for Forest Inventory Utilizing
Three Sources of Information.''} \emph{Annals of Forest Science} 73 (4):
895--910.

\leavevmode\vadjust pre{\hypertarget{ref-sarndal2003model}{}}%
Särndal, Carl-Erik, Bengt Swensson, and Jan Wretman. 2003. \emph{Model
Assisted Survey Sampling}. Springer Science \& Business Media.

\leavevmode\vadjust pre{\hypertarget{ref-shoot2021foresttypeclass}{}}%
Shoot, Caileigh, Hans-Erik Andersen, L. Monika Moskal, Chad Babcock,
Bruce D. Cook, and Douglas C. Morton. 2021. {``Classifying Forest Type
in the National Forest Inventory Context with Airborne Hyperspectral and
Lidar Data.''} \emph{Remote Sensing} 13 (10).
\url{https://doi.org/10.3390/rs13101863}.

\leavevmode\vadjust pre{\hypertarget{ref-staahl2011model}{}}%
Ståhl, Göran, Sören Holm, Timothy G Gregoire, Terje Gobakken, Erik
Næsset, and Ross Nelson. 2011. {``Model-Based Inference for Biomass
Estimation in a LiDAR Sample Survey in Hedmark County, Norway.''}
\emph{Canadian Journal of Forest Research} 41 (1): 96--107.

\leavevmode\vadjust pre{\hypertarget{ref-staahl2016use}{}}%
Ståhl, Göran, Svetlana Saarela, Sebastian Schnell, Sören Holm, Johannes
Breidenbach, Sean P Healey, Paul L Patterson, et al. 2016. {``Use of
Models in Large-Area Forest Surveys: Comparing Model-Assisted,
Model-Based and Hybrid Estimation.''} \emph{Forest Ecosystems} 3 (1):
1--11.

\leavevmode\vadjust pre{\hypertarget{ref-strimburepeatlidarsampling2017}{}}%
Strîmbu, Victor Felix, Liviu Theodor Ene, Terje Gobakken, Timothy G.
Gregoire, Rasmus Astrup, and Erik Næsset. 2017. {``Post-Stratified
Change Estimation for Large-Area Forest Biomass Using Repeated ALS Strip
Sampling.''} \emph{Canadian Journal of Forest Research} 47 (6): 839--47.

\leavevmode\vadjust pre{\hypertarget{ref-strunk2014prediction}{}}%
Strunk, Jacob L, Hailemariam Temesgen, Hans-Erik Andersen, and Petteri
Packalen. 2014. {``Prediction of Forest Attributes with Field Plots,
Landsat, and a Sample of Lidar Strips.''} \emph{Photogrammetric
Engineering \& Remote Sensing} 80 (2): 143--50.

\leavevmode\vadjust pre{\hypertarget{ref-usda2016forest}{}}%
USDA Forest Service. 2016. \emph{Forest Inventory and Analysis Strategic
Plan}. FS-1079. USDA Forest Service Washington, DC, USA.

\leavevmode\vadjust pre{\hypertarget{ref-VandenBoogaart2013CompositionalR}{}}%
Van den Boogaart, K Gerald, and Raimon Tolosana-Delgado. 2013.
\emph{{Analyzing compositional data with R}}. Berlin: Springer.
\url{https://doi.org/10.1007/978-3-642-36809-7}.

\leavevmode\vadjust pre{\hypertarget{ref-westfallgreenbooktwo2022}{}}%
Westfall, James A., John W. Coulston, Gretchen G. Moisen, and Hans-Erik
Andersen, eds. 2022. \emph{Sampling and Estimation Documentation for the
Enhanced Forest Inventory and Analysis Program}. NRS-GTR-207. USDA
Forest Service, Northern Research Station.
\url{https://doi.org/10.2737/nrs-gtr-207}.

\end{CSLReferences}

\begin{appendices}
\section{Variance of ratio-of-ratios estimator of the domain density}

\setcounter{equation}{0} 
\renewcommand\theequation{A.\arabic{equation}} 

In deriving the variance estimation for the $RoR$ estimator of domain density without poststratification, we note that a first order Taylor approximation leads to

\begin{equation}
V\left( {\widehat{D}}_{RoR} \right) = \ V\left\lbrack \frac{\sum_{i = 1}^{m}{\ {\widehat{t}}_{ri}}}{\sum_{i = 1}^{m}{\widehat{a}}_{ri}}\  \right\rbrack\  \approx \ \frac{1}{E\left(\left( \sum_{i = 1}^{m}{\widehat{a}}_{i} \right)^{2}\right)}\ \ V\left\lbrack \sum_{i = 1}^{m}\left( {\widehat{t}}_{ri} - D{\widehat{a}}_{ri} \right) \right\rbrack = \frac{1}{E\left(\left( \sum_{i = 1}^{m}{\widehat{a}}_{i} \right)^{2}\right)} V\left( \cdot \right)
\end{equation} 

Conditioning of the first stage sample leads to: \( V\left( \cdot \right) = V_{1}\left(E_{2}\left( \cdot \right)\right) + E_{1}\left(V_{2}\left( \cdot \right)\right) \). This means that we separate the variance into $(1)$ a first term which is the variance, among first stage sampling units, of their expected values from second stage sampling and $(2)$ a second term which is the variance from second stage sampling. This is a common technique applied for deriving variances following multistage or multiphase sampling (e.g., Cochran (1977)). We start by looking at \( V\left( \cdot \right) = V_{1}\left(E_{2}\left( \cdot \right)\right) \), that is \( V_{1}\left(E_{2}\left( \sum_{i = 1}^{m} {\widehat{t}}_{ri} - D{\widehat{a}}_{ri} \right) \right) \). Because the second stage expectations are (at least approximately) the true values of $t$ and $a$ for all first stage units, we obtain \( V_{1}\left(\sum_{i = 1}^{m} {\widehat{t}}_{ri} - D{\widehat{a}}_{ri} \right) = m(1 - \frac{m}{M}) S_{u}^{2} = m(\frac{1}{m} - \frac{1}{M}) S_{u}^{2} \) where $D = D_{RoR}$ and $S_{u}^{2} = \frac{1}{M - 1}\sum_{i = 1}^{M}{({u}_{i} - \overline{u})}^{2}$. 

Thus 
\begin{multline} 
V_{1}\left(E_{2}\left( {\widehat{D}}_{RoR} \right)\right) = \frac{1}{E\left(\left( \sum_{i = 1}^{m}{\widehat{a}}_{i} \right)^{2}\right)} m^{2} (\frac{1}{m} - \frac{1}{M}) S_{u}^{2} \\ 
= \frac{1}{\frac{M^{2}}{m^{2}}E\left(\left( \sum_{i = 1}^{m}{\widehat{a}}_{i} \right)^{2}\right)} M^{2} (\frac{1}{m} - \frac{1}{M}) S_{u}^{2} \\ 
=  \frac{1}{E\left( \widehat{A^*}^{2} \right)} M^{2} (\frac{1}{m} - \frac{1}{M}) S_{u}^{2} \\
\end{multline}

This is the first term of the variance. An estimator is obtained by substituting \( E\left( \widehat{A^*}^{2} \right) \) by \( \widehat{A^*}^{2} \) and  
\( S_{u}^{2} \) by \( s_{u}^{2} \), i.e. the estimator based on the sample. 

We then turn to the second term of the Taylor approximation, i.e. \( E_{1}\left(V_{2}\left( \cdot \right)\right) \), which is the expectation across first stage units of the second stage variances. Thus, \( E_{1}\left(V_{2}\left( \sum_{i = 1}^{m} {\widehat{t}}_{ri} - D{\widehat{a}}_{ri} \right) \right) \). Variability in the second stage is due to subsampling field plots, where 

\begin{multline}
V_{2}\left(\sum \widehat{t}_{ri} - D{\widehat{a}}_{ri} \right) \\
= V_{2} \left(\sum \left(\sum{C_{1} + \frac{N_{i}}{n_{i}} \sum e_{ik}^{y}}\right)  - D\left(\sum{C_{2} + \frac{N_{i}}{n_{i}} \sum e_{ik}^{a}}\right) \right) \\
= V_{2} \left(\sum_{}^{m} \left(\frac{N_{i}}{n_{i}}\left( \sum \left( e_{ik}^{y} - De_{ik}^{a} \right)  \right) \right) \right) \\
= V_{2} \left(\sum_{}^{m} \left(\frac{N_{i}}{n_{i}}\left( \sum v_{ik}  \right) \right) \right)  \\
= \sum_{}^{m} \frac{N_{i}^{2}}{n_{i}^{2}} V_{2} \left( \sum_{k}^{} v_{ik} \right) \\
= \sum_{}^{m} \frac{N_{i}^{2}}{n_{i}^{2}} n_{i} \left( 1 - \frac{n_{i}}{N_{i}} \right) S_{vi}^{2} \\
= \sum_{}^{m} N_{i}^{2} \left( \frac{1}{n_{i}} - \frac{1}{N_{i}} \right) S_{vi}^{2}  \\
\end{multline}

Where $v_{ik} = e_{ik}^{y} - De_{ik}^{a}$, $C_{1}$ are predictions of $y_{i}$ (with error terms $e_{ik}^{y}$), $C_{2}$ are predictions of $a_{i}$ (with error terms $e_{ik}^{a}$), and $C_{1}$ and $C_{2}$ are constants that do not affect the variance. 

Next, we should take the expectation over first stage samples:

\begin{multline}
E_{1}\left(\sum_{}^{m} N_{i}^{2}\left(\frac{1}{n_{i}} - \frac{1}{N_{i}}\right) S_{vi}^{2} \right)  \\
= \sum_{}^{M} \pi_{i}N_{i}^{2}\left(\frac{1}{n_{i}} - \frac{1}{N_{i}}\right) S_{vi}^{2} \\
= \sum_{}^{M} \frac{m}{M}N_{i}^{2}\left(\frac{1}{n_{i}} - \frac{1}{N_{i}}\right) S_{vi}^{2} \\
= \frac{m}{M}\sum_{}^{M}N_{i}^{2}\left(\frac{1}{n_{i}} - \frac{1}{N_{i}}\right) S_{vi}^{2} \\
\end{multline}

where $\pi_{i}$ is the inclusion probability of the $i^{th}$ \emph{PSU}. 

Lastly, we should multiply this term by \(  \frac{1}{{E\left( \sum_{}^{m} {\widehat{a}}_{ri} \right)}^{2}} \) leading to:

\begin{multline}
E_{1}V_{2}\left({\widehat{D}}_{RoR}\right) =  \frac{1}{{E\left( \sum_{}^{m} {\widehat{a}}_{ri} \right)}^{2}} \frac{m}{M}\sum_{}^{M}N_{i}^{2}\left(\frac{1}{n_{i}} - \frac{1}{N_{i}}\right) S_{vi}^{2} \\
=  \frac{1}{{\frac{M^{2}}{m^{2}}E\left( \sum_{}^{m} {\widehat{a}}_{ri} \right)}^{2}} \frac{M}{m}\sum_{}^{M}N_{i}^{2}\left(\frac{1}{n_{i}} - \frac{1}{N_{i}}\right) S_{vi}^{2} \\
=  \frac{1}{{E\left( \widehat{A^*}\right)}^{2}} \frac{M}{m}\sum_{}^{M}N_{i}^{2}\left(\frac{1}{n_{i}} - \frac{1}{N_{i}}\right) S_{vi}^{2} \\
\end{multline}

This is the second term in the variance. To obtain an estimator, we substitute \( {E\left( \widehat{A^*}\right)}^{2} \) by \( \widehat{A^*}^{2} \) and  
\( S_{vi}^{2} \) by \( s_{vi}^{2} \) for the sampled $PSU$s. Also, we change the summation from $M$ to $m$. (This might appear to lead to a underestimation of the variance, but this underestimation if compensated by using estimated rather than the true values when \( S_{u}^{2} \) is estimated (e.g., see Cochran (1977) or  Särndal, Swensson, and Wretman (2003))). The resulting variance estimator is therefore: 

\begin{equation}
\widehat{V}\left( {\widehat{D}}_{RoR} \right) = \frac{1}{\widehat{A^*}^{2}}\left\lbrack M^{2}\left( \frac{1}{m} - \frac{1}{M} \right)s_{u}^{2} + \frac{M}{m}\sum_{S_{1}}^{}N_{i}^{2}\left( \frac{1}{n_{i}} - \frac{1}{N_{i}} \right)s_{vi}^{2} \right\rbrack
\end{equation}

\section{Variance of ratio-of-ratios estimator of the domain density with poststratification}

\setcounter{equation}{0} 
\renewcommand\theequation{B.\arabic{equation}}

\begin{multline}
V\left( {\widehat{D}}_{RoR,PS} \right) \approx \frac{1}{{E\left( {\widehat{A}}_{R,PS}\right)}^{2}}V\left( {\widehat{t}}_{R,PS} - D_{RoR,PS}{\widehat{A}}_{R,PS} \right) = \\ 
\frac{1}{{E\left( {\widehat{A}}_{R,PS}\right)}^{2}}\left\lbrack V\left( {\widehat{t}}_{R,PS}) + D_{RoR,PS}^{2}{V(\widehat{A}}_{R,PS}) \right) -  
2D_{RoR,PS}Cov({\widehat{t}}_{R,PS},{\widehat{A}}_{R,PS}) \right\rbrack
\end{multline}

We expand each of the terms inside the square brackets,
reflecting the two-stage sampling procedure, i.e.
\(V\left( {\widehat{t}}_{R,PS} \right) = V_{1}(E_{2}\left( {\widehat{t}}_{R,PS} \right)) + E_{1}(V_{2}({\widehat{t}}_{R,PS})\),
and similarly for \({V(\widehat{A}}_{R,PS})\ \)and
\(Cov({\widehat{t}}_{R,PS},{\widehat{A}}_{R,PS})\). In the following this
will be done for \(V\left( {\widehat{t}}_{R,PS} \right)\) and
\(Cov({\widehat{t}}_{R,PS},{\widehat{A}}_{R,PS})\), noting that the formula
for \({V(\widehat{A}}_{R,PS})\) will be similar to the formula for
\(V\left( {\widehat{t}}_{R,PS} \right)\). Thus,

\begin{equation}
V\left( {\widehat{t}}_{R,PS} \right) = V\left( \sum_{h}^{}{\widehat{t}}_{Rh} \right) = \sum_{h}^{}{V(N_{h}\frac{\sum_{i = 1}^{m}{\ {\widehat{t}}_{rhi}}}{\sum_{i = 1}^{m}N_{hi}}}) + \sum_{h}^{}{\sum_{g \neq h}^{}{Cov(}}N_{h}\frac{\sum_{i = 1}^{m}{\ {\widehat{t}}_{rhi}}}{\sum_{i = 1}^{m}N_{hi}},N_{g}\frac{\sum_{i = 1}^{m}{\ {\widehat{t}}_{rgi}}}{\sum_{i = 1}^{m}N_{gi}})
\end{equation}

In (B.2), each of the terms
\(V(N_{h}\frac{\sum_{i = 1}^{m}{\ {\widehat{t}}_{rhi}}}{\sum_{i = 1}^{m}N_{hi}})\)
can be further developed as

\begin{equation}
V\left( N_{h}\frac{\sum_{i = 1}^{m}{\ {\widehat{t}}_{rhi}}}{\sum_{i = 1}^{m}N_{hi}} \right) \approx \frac{N_{h}^{2}}{{(E\sum_{i = 1}^{m}{N_{hi})}}^{2}}V\left( \sum_{i = 1}^{m}{\ ({\widehat{t}}_{rhi}} - R_{h}N_{hi}) \right)
\end{equation}

In (B.3), the \(R_{h}\)-terms are the stratum-specific ratios
\(\frac{\sum_{i = 1}^{m}{\ t_{rhi}}}{\sum_{i = 1}^{m}N_{hi}}\).
Addressing
\(V\left( \sum_{i = 1}^{m}{\ ({\widehat{t}}_{rhi}} - R_{h}N_{hi}) \right)\)
through \(V_{1}(E_{2}(.)) + E_{1}(V_{2}(.))\), the first term will be
\(V\left( \sum_{i = 1}^{m}{\ (t_{rhi}} - R_{h}N_{hi}) \right)\) and
the second term
\(\frac{m}{M}V\left( \sum_{i = 1}^{M}{\ ({\widehat{t}}_{rhi}}) \right)\).
Again, we use \(r_{h} = t_{rhi} - R_{h}N_{hi}\). Using simple
random sampling without replacement the variance of the first term is
\(m^{2}\left( \frac{1}{m} - \frac{1}{M} \right)S_{rh}^{2}\) and the
variance of the second term
\(\frac{1}{M^{2}}\sum_{M}^{}N_{hi}^{2}(\frac{1}{n_{hi}} - \frac{1}{N_{hi}})\ S_{e,h}^{2}\).
Here,
\(S_{e,h}^{2} = \frac{1}{N_{hi} - 1}\sum_{k = 1}^{N_{hi}}\left( e_{k,h} - {\overline{e}}_{hi} \right)^{2}\)
with \({\overline{e}}_{hi} = \frac{\sum_{}^{}e_{k,h}}{N_{hi}}\). Thus,
each of the variance terms in (B.3) can be expressed as

\begin{multline}
V\left( N_{h}\frac{\sum_{i = 1}^{m}{\ {\widehat{t}}_{rhi}}}{\sum_{i = 1}^{m}N_{hi}} \right) = \\
\frac{N_{h}^{2}}{{(\frac{m}{M}\sum_{i = 1}^{M}{N_{hi})}}^{2}}\left\lbrack m^{2}\left( \frac{1}{m} - \frac{1}{M} \right)S_{rh}^{2} + \frac{m^{2}}{M^{2}}\sum_{M}^{}N_{hi}^{2}\left( \frac{1}{n_{hi}} - \frac{1}{N_{hi}} \right)S_{e,h}^{2} \right\rbrack = \\
\frac{N_{h}^{2}}{{(\frac{1}{M}\sum_{i = 1}^{M}{N_{hi})}}^{2}}\left\lbrack \left( \frac{1}{m} - \frac{1}{M} \right)S_{rh}^{2} + \frac{1}{M^{2}}\sum_{M}^{}N_{hi}^{2}\left( \frac{1}{n_{hi}} - \frac{1}{N_{hi}} \right)S_{e,h}^{2} \right\rbrack =  \\
\frac{N_{h}^{2}}{{\overline{N}}_{h}^{2}}\left\lbrack M^{2}\left( \frac{1}{m} - \frac{1}{M} \right)S_{rh}^{2} + \sum_{M}^{}N_{hi}^{2}\left( \frac{1}{n_{hi}} - \frac{1}{N_{hi}} \right)S_{e,h}^{2} \right\rbrack
\end{multline} where ${\overline{N}}_{h}$ is the average across all strips.

\(\ \)A variance estimator corresponding to the variance in (B.4) would
be

\begin{equation}
\widehat{V}\left( N_{h}\frac{\sum_{i = 1}^{m}{\ {\widehat{t}}_{rhi}}}{\sum_{i = 1}^{m}N_{hi}} \right) = \ \frac{N_{h}^{2}}{{\widehat{N}}_{h}^{2}}\left\lbrack M^{2}\left( \frac{1}{m} - \frac{1}{M} \right)s_{rh}^{2} + \frac{M}{m}\sum_{S1}^{}N_{hi}^{2}\left( \frac{1}{n_{hi}} - \frac{1}{N_{hi}} \right)s_{e,h}^{2} \right\rbrack
\end{equation}

The variance \({V(\widehat{A}}_{R,PS})\ \)in (B.1) and its estimator would
be similar, using ``area'' instead of biomass in \(s_{rh}^{2}\) and
\(s_{e,h}^{2}\).

Next, we address the covariance terms in (B.2), i.e.
\(Cov(N_{h}\frac{\sum_{i = 1}^{m}{\ {\widehat{t}}_{rhi}}}{\sum_{i = 1}^{m}N_{hi}},N_{g}\frac{\sum_{i = 1}^{m}{\ {\widehat{t}}_{rgi}}}{\sum_{i = 1}^{m}N_{gi}})\).
The second term of \(Cov(.,.) = {Cov}_{1}(E_{2}(.,.)) + E_{1}({Cov}_{2}(.,.)\) should be
zero, since no covariance should arise due to the second stage
subsampling. Thus, the covariance would be limited to the first term,
i.e. to
\(Cov(N_{h}\frac{\sum_{i = 1}^{m}{\ t_{rhi}}}{\sum_{i = 1}^{m}N_{hi}},N_{g}\frac{\sum_{i = 1}^{m}{\ t_{rgi}}}{\sum_{i = 1}^{m}N_{gi}})\).
Using Taylor approximation this covariance can be re-written as

\begin{multline}
Cov\left( N_{h}\frac{\sum_{i = 1}^{m}{\ t_{rhi}}}{\sum_{i = 1}^{m}N_{hi}},N_{g}\frac{\sum_{i = 1}^{m}{\ t_{rgi}}}{\sum_{i = 1}^{m}N_{gi}} \right) \approx \\ \frac{N_{h}N_{g}}{E(\sum_{i}^{m}{N_{hi})E(\sum_{i}^{m}N_{gi})}}Cov\left( \sum_{i = 1}^{m}{t_{rhi}} - R_{h}N_{hi},\sum_{i = 1}^{m}{\ t_{irg}} - R_{g}N_{ig} \right) = \\ \frac{N_{h}N_{g}}{E(\sum_{i}^{m}{N_{hi})E(\sum_{i}^{m}{N_{gi})}}}Cov\left( \sum_{i = 1}^{m}{r_{hi}},\sum_{i = 1}^{m}{r_{gi}} \right)
\end{multline}

Thus, under simple random sampling without replacement the covariance
would be

\begin{multline}
Cov\left( N_{h}\frac{\sum_{i = 1}^{m}{\ t_{rhi}}}{\sum_{i = 1}^{m}N_{hi}},N_{g}\frac{\sum_{i = 1}^{m}{\ t_{rgi}}}{\sum_{i = 1}^{m}N_{gi}} \right) = \\
 \frac{N_{h}N_{g}}{E(\sum_{i}^{m}{N_{hi})E(\sum_{i}^{m}{N_{gi})}}}m\ \left( 1 - \frac{m}{M} \right)Cov\left( r_{h},r_{g} \right) = \\ \frac{N_{h}N_{g}}{E\left( {\widehat{N}}_{h} \right)E({\widehat{N}}_{g})}M^{2}\left( 1 - \frac{m}{M} \right)\frac{Cov\left( r_{h},r_{g} \right)}{m}
\end{multline}

Again, a covariance estimator would use the estimates of \(E( {\widehat{N}}_{h})\) and \(E({\widehat{N}}_{g})\) and \(Cov\left( r_{h},r_{g} \right)\) estimated in the usual way across the $m$ survey strips. The covariance for the terms related to \({\widehat{A}}_{R,PS}\) should be computed in a similar way, with
``area'' inputs rather than biomass inputs.

What remains from equation (B.1) is the term
\(Cov\left( {\widehat{t}}_{R,PS},{\widehat{A}}_{R,PS} \right) = Cov\left( \sum_{h}^{}{\widehat{t}}_{Rh},\sum_{h}^{}{\widehat{A}}_{Rh} \right) = \sum_{h}^{}{\sum_{g}^{}{{Cov(\widehat{t}}_{Rh},{\widehat{A}}_{Rg})}}\).
Like before, it is not likely that the second stage subsampling leads to
covariance and thus only the covariance conditional on the stage 2
expectations should be addressed. This leads to

\begin{multline}
Cov\left( N_{h}\frac{\sum_{i = 1}^{m}{\ {\widehat{t}}_{rhi}}}{\sum_{i = 1}^{m}N_{hi}},N_{g}\frac{\sum_{i = 1}^{m}{\ {\widehat{a}}_{rgi}}}{\sum_{i = 1}^{m}N_{gi}} \right) \approx \\ \frac{N_{h}N_{g}}{E(\sum_{i}^{m}{N_{hi})E(\sum_{i}^{m}N_{gi})}}Cov\left( \sum_{i = 1}^{m}{\ t_{rhi}} - R_{h}N_{hi},\sum_{i = 1}^{m}{\ a_{rgi}} - R_{a,g}N_{gi} \right) = \\ \frac{N_{h}N_{g}}{E(\sum_{i}^{m}{N_{hi})E(\sum_{i}^{m}N_{gi})}}Cov\left( \sum_{i = 1}^{m}{\ r_{hi}},\sum_{i = 1}^{m}{\ r_{a,gi}} \right) = \\ \frac{N_{h}N_{g}}{E\left( {\widehat{N}}_{h} \right)E({\widehat{N}}_{g})}M^{2}\left( 1 - \frac{m}{M} \right)\frac{Cov\left( r_{h},r_{a,g} \right)}{m}
\end{multline}

An estimator for \({V(\widehat{A}}_{R,PS})\) would be derived similarly. 

\end{appendices}

\end{document}